\documentclass{article}
\usepackage{a4wide}
\usepackage{tabularx}
\usepackage{adjustbox}
\usepackage{multirow}
\usepackage{graphicx}
\usepackage[numbers]{natbib}
\usepackage{amsfonts}
\usepackage{booktabs}
\usepackage{amsmath,amssymb,amsthm,bm}
\usepackage{hyperref}
\usepackage{calc}
\usepackage[labelfont=bf]{caption}
\usepackage{float}
\usepackage[final]{changes}

\title{Dynamic functional brain connectivity results depend on modeling assumptions: comparing the sliding-window method and the Wishart process for dynamic hypothesis testing}
\author{\small Hester Huijsdens$^1$, Linda Geerligs$^1$, Max Hinne$^1$ \\ \\ \small $^1$ Donders Institute, Radboud University, Nijmegen, The Netherlands}
\date{}

\newcommand{\Observations}{N}
\newcommand{\Regions}{D}
\newcommand{\Subjects}{S}
\newcommand{\windowlength}{\lambda}
\newcommand{\Dof}{K}
\renewcommand{\v}[1]{\boldsymbol{\mathbf{#1}}}

\begin{document}

\maketitle
\begin{abstract}
Understanding the temporal dynamics of functional brain connectivity is important for addressing various questions in network neuroscience, such as how connectivity affects cognition and changes with disease. A fundamental challenge is to evaluate whether connectivity truly exhibits dynamics, or simply is static. The most common approach uses sliding-window methods to model functional connectivity over time, and this is often combined with frequentist hypothesis testing frameworks to evaluate dynamics. However, this requires defining appropriate sampling distributions and hyperparameters, such as window length, which imposes specific assumptions on the dynamics. Here, we explore how these assumptions influence the detection of dynamic connectivity, and introduce an alternative approach based on Bayesian hypothesis testing with Wishart processes. This framework estimates uncertainty in the connectivity estimates, and uses this to provide strength of evidence for both dynamic and static connectivity. It encodes assumptions through prior distributions, allowing prior knowledge on the time-dependent structure of connectivity to be incorporated into the model. Using simulations, we compare the two approaches and demonstrate how different assumptions affect the detection of dynamic connectivity. Finally, by applying both approaches to an fMRI working-memory task, we find that conclusions at the group-level increase robustness to modeling choices. Our work highlights the importance of carefully considering modeling assumptions when evaluating dynamic connectivity.
\end{abstract}

\section{Introduction}
Dynamic (time-varying) functional connectivity is explored in an increasing number of studies in network neuroscience~\cite{kampman2024time, liegeois2019resting, lurie2020questions, alonso2025can}. Unlike traditional functional connectivity analyses, in which a single estimate of connectivity is estimated from time-series data, dynamic functional connectivity aims to capture how the brain's functional connectivity changes over time. These dynamic estimates have been shown to serve as more sensitive biomarkers than single time-averaged estimates in different domains~\cite{lurie2020questions, calhoun2014chronnectome}. For example, functional connectivity dynamics are altered during aging~\cite{battaglia2020dynamic}, and can improve the classification of patients diagnosed with neurological disorders and healthy controls~\cite{saha2021classification, sakouglu2010method}. 

Several different approaches exist to estimate dynamic connectivity from time series data, such as multivariate  autoregressive models~\cite{lindquist2014evaluating, silvennoinen2009multivariate} and hidden Markov models (HMMs)~\cite{shappell2019improved, vidaurre2017brain}. By far the most popular method is the sliding-window approach~\cite{sakouglu2010method, allen2014tracking}. Although these procedures have produced promising findings, there are ongoing discussions related to the interpretation of this dynamic functional connectivity~\cite{lurie2020questions}. For instance, studies have demonstrated that the observed fluctuations might not reflect just true neural dynamics, but can instead be attributed to sampling variability, head motion artifacts ~\cite{laumann2017stability}, and modeling choices~\cite{hindriks2016can}. For example, when using a sliding-window approach~\cite{sakouglu2010method, allen2014tracking}, the choice of window length greatly influences the estimates of dynamic connectivity. Therefore, an important question is whether perceived dynamics are \emph{statistically meaningful} and truly reflect dynamics, instead of artifacts caused by the estimation procedure. 

The solution to this question is straightforward: perform a statistical test to distinguish signal (a truly dynamic connection) from noise. Typically, these tests are based on frequentist null hypothesis testing~\cite{hindriks2016can, sakouglu2010method}, and proceed as follows. First, a null distribution is generated that preserves the important characteristics of the observed data, while removing any meaningful dynamics. Any remaining dynamics in the null distribution are random and reflect properties of the observed data, such as autocorrelation in the time series itself. Next, a dynamic connectivity estimation method (such as a sliding-window method) is applied to obtain estimates of dynamic functional connectivity for both the null distribution and the observed data. Finally, to evaluate whether the observed connectivity is static or dynamic, a test statistic is computed over the connectivity estimates of the observed and null data. These values are then compared, resulting in a $p$-value describing how (un)likely it is to see these observations if the null hypothesis was true. 

One can also perform the entire procedure -- from estimating dynamic connectivity to statistical testing -- using the Bayesian framework. Although some Bayesian approaches for estimating dynamic connectivity have been proposed, such as methods using Bayesian HMMs~\cite{warnick2018bayesian, taghia2017bayesian}, Bayesian MGARCH~\cite{virbickaite2015bayesian} and Bayesian nonparametric models~\cite{wilson2010generalised, kampman2024time}, few studies have focused on Bayesian \emph{testing for dynamics}. However, Bayesian testing potentially offers several benefits. For instance, Bayesian approaches explicitly quantify estimation uncertainty, both over the parameters of the estimation approaches, as well as over the hypotheses. This allows a Bayesian test to find evidence in favor of the null hypothesis (indicating that the functional connectivity is static), which is not possible under the frequentist approach~\cite{rouder_bayesian_2009}.

Both the frequentist and the Bayesian approach lead to an answer as to whether a connection is truly dynamic. However, in the steps leading to the statistical test there are several modeling choices that affect the final conclusion. In this paper, we investigate the impact of typical assumptions in these pipelines, such as the size of the sliding-window, or the expected kind of autocorrelations in a Bayesian model. While the combination of sliding-window estimation with frequentist testing is widely established, Bayesian alternatives are much less common. Here, we adopt a recently introduced Bayesian nonparametric approach, known as the Wishart process~\cite{wilson2010generalised,kampman2024time,huijsdens2024robust}. This approach is particularly suited for testing the impact of modeling assumptions, as different expectations on the kind of dynamics can be expressed explicitly by selecting the corresponding covariance function~\cite{kampman2024time,huijsdens2024robust}. In addition, we introduce a Bayesian hypothesis test using the Wishart process estimate.

The paper is organized as follows. In Section~\ref{section:methods}, we first describe the sliding-window method and the Wishart process. We then recap the traditional frequentist test for dynamic connectivity, and introduce a Bayesian statistical test for dynamic functional connectivity. In Section~\ref{section:simulation_results}, we compare the two frameworks using simulation studies and show how, with the Bayesian Wishart process method, uncertainty quantification can provide more information about dynamic functional connectivity. Moreover, in Section~\ref{section:empirical_results}, we illustrate the use of both frameworks on a task-based fMRI data from the Human Connectome Project~\cite{van2013wu}. Finally, in Section~\ref{section:discussion}, we conclude our comparison and discuss future directions of research.

\section{Methods} \label{section:methods}
Before we proceed with a discussion of the two statistical frameworks and dynamic connectivity methods, we present some notation and terminology that will be used throughout the rest of this paper. We use upper case letters $X$ for scalar values, bold lower case letters $\v{x}$ for vectors, and bold upper case letters $\v{X}$ for matrices. The time series data consists of $\Observations$ observations over $\Regions$ variables (i.e. brain regions), with each observation denoted by $\v{y}_n \in \mathbb{R}^\Regions$. Stacked together, these form the matrix $\v{Y} = \left(\v{y}_1, \ldots, \v{y}_\Observations \right)^\top \in \mathbb{R}^{\Observations \times \Regions}$. 

\subsection{Hypothesis testing for dynamic connectivity}
Identifying whether one or more pairs of regions are truly dynamically connected, requires the following steps:
\begin{enumerate}
    \item First, we must estimate the functional connectivity over time itself, as this is not directly observed. This is commonly done using a form of the sliding-window method~\cite{hindriks2016can,Preti2017a}, which we discuss in Section~\ref{subsubsection:sliding_window}. Alternatively, one can use a Bayesian approach, for example based on Wishart processes~\cite{kampman2024time, huijsdens2024robust}, which are described in sections~\ref{subsubsection:wishart_process} and \ref{subsubsection:pooled_gwp}.
    \item To test for meaningful dynamics in connectivity, the connections are typically summarized using test statistics. These are presented in Section~\ref{subsection:test_statistics}.
    \item Lastly, the test statistics are used for hypothesis testing, which can be done in either the frequentist, $p$-value based framework, or in a Bayesian way, which we introduce in this work. The two approaches for testing are described in Section~\ref{subsection:testing}.
\end{enumerate} 

\subsection{Dynamic connectivity estimation} \label{subsection:dynamic_connectivity}
We are interested in the dynamic connectivity between all pairs of brain regions. Throughout this paper, we use the following notation to refer to (dynamic) connectivity estimates. We use $\v\Sigma \in \mathbb{R}^{\Regions \times \Regions}$ to refer to a single (positive-definite) covariance matrix, such as obtained by a sliding-window estimate. In addition, $\v\Sigma(x_n) \in \mathbb{R}^{\Regions \times \Regions}$ refers to the covariance specifically at input location $x_n$. Finally, $\v\Sigma(\v x) \in \mathbb{R}^{\Observations \times \Regions \times \Regions}$ represents a sequence of covariance matrices, such that $\v\Sigma(\v x) = \left(\v\Sigma(x_1), \ldots, \v\Sigma(x_\Observations)\right)$.

\subsubsection{Sliding-window methods} \label{subsubsection:sliding_window}
The intuition behind the sliding-window approach is simple: rather than computing a single covariance matrix of the observations $\v{Y}$ using $\v{\Sigma} = \frac{1}{\Observations-1} \v{Y}^\top \v{Y}$, we compute $\v{\Sigma}$ for a series of consecutive segments of $\v{Y}$ instead~\cite{sakouglu2010method, allen2014tracking, hindriks2016can}. The length of this segment, $\windowlength$, is known as the window size, and determines how many observations are used for each covariance estimate. To get an estimate of covariance over time, the window is shifted by a number of observations, called the stride length $\tau$. 

The method is described in detail in Supplementary Section~\ref{supplementary:sliding_window}. Although this approach is easy to implement, it has an important limitation. Namely, both the size of the window and the stride length must be determined by the practitioner, while it has been shown that this parameter has a substantial effect on the estimated connectivity~\cite{leonardi2015spurious, shakil2016evaluation, lamovs2018spatial}. A small window length will result in rapidly fluctuating estimates of connectivity, whereas with a large window size, relevant fluctuations in the connectivity might be smoothed out.

\subsubsection{Wishart processes} \label{subsubsection:wishart_process}
We propose to combine the Bayesian framework for hypothesis testing with Wishart processes~\cite{heaukulani2019scalable, wilson2010generalised}, a Bayesian model that was introduced by Wilson and Ghahramani~\cite{wilson2010generalised} and has later been used to model dynamic covariance in different domains such as neuroscience and psychology~\cite{kampman2024time, huijsdens2024robust, meng2023dynamic}. Wishart processes allow us to set prior assumptions on different aspects of the connectivity, such as on the type of connectivity structure over time.

To explain the Wishart process, it is helpful to first look at how to estimate a single $\Regions \times \Regions$ covariance matrix $\v{\Sigma}$, as in the static connectivity case, using a Bayesian model. A straightforward Bayesian model for $\v{\Sigma}$ is:
\begin{equation} \label{eq:sigma_and_y_mvn}
    \begin{split}
        \v\Sigma &\sim \mathcal{W}(\v V, \Dof) \\
        \v{y}_n &\sim \mathcal{MVN}_\Regions(\v 0, \v\Sigma) \enspace, \quad n = 1, \ldots, \Observations \enspace,
    \end{split}
\end{equation} 
in which $\mathcal{MVN}_\Regions(\v 0, \v\Sigma)$ is the $\Regions$-dimensional multivariate Gaussian distribution with mean $\v 0$ and covariance matrix $\v\Sigma$, and $\mathcal{W}(\v V, \Dof)$ is the Wishart \emph{distribution} with scale matrix $\v V$ and $\Dof$ degrees of freedom. As the Wishart distribution is conjugate to the multivariate Gaussian, the posterior distribution $p(\v \Sigma \mid \v{Y})$ can be computed analytically~\cite{zhang2021note}. 

Another way of looking at the Wishart distribution is by how samples from this distribution are generated. Using again the multivariate Gaussian distribution, a Wishart-distributed covariance matrix can be obtained using:
\begin{equation}\label{eq:wishart_distribution}
    \begin{split}
        \v f_k &\sim \mathcal{MVN}_\Regions(\v{0}, \v I) \enspace, \quad k = 1, \ldots, \Dof \enspace \\
        \v \Sigma &= \sum_{k=1}^\Dof \v{L}\v{f}_k\v{f}_k^\top \v{L}^\top \sim \mathcal{W}(\v V, \Dof) \enspace,
    \end{split}
\end{equation} in which $\v V = \v{L}\v{L}^\top$ is the Cholesky decomposition of the scale matrix $\v V$. In words, the sum of outer products of $\Dof$ i.i.d. zero-mean multivariate Gaussian samples is Wishart distributed~\cite{ghosh2002simple}.
 
Extending this idea, the Wishart \emph{process} instead models a covariance matrix that changes with input. The Wishart process is constructed similarly to Eq.~\eqref{eq:wishart_distribution}, but the multivariate Gaussian variates are now replaced by independent zero-mean Gaussian processes (GPs)~\cite{rasmussen2005}: 
\begin{equation} \label{eq:f_samples}
    \begin{split}
        f_{ki} &\sim \mathcal{GP} \left( 0, \kappa(\cdot, \cdot; \theta) \right) \enspace, \quad k = 1, \ldots, \Dof \enspace, \quad i = 1, \ldots, \Regions \enspace,
    \end{split}
\end{equation} where the mean function is set to \ensuremath{0} and $\kappa(\cdot, \cdot;\theta)$ is the covariance function with parameters $\theta$. We collect the evaluations of these functions as $\v{f}_k \left( x_n \right) = \left( f_{k1} \left( x_n \right), \ldots, f_{k\Regions} \left( x_n \right) \right)^\top$. The covariance function determines several properties, such as smoothness, of the Wishart process. Moreover, if we assume that $\kappa \left( x_n, x_n ;\theta \right) = 1$ for all observations $n = 1, \ldots, \Observations$, we can separate the correlations from the scale matrix $\v{V}$ when constructing the Wishart process:
\begin{equation} \label{eq:wishart_process}
    \v{\Sigma} \left( x_n \right) = \sum_{k=1}^\Dof \v{L} \v{f}_k \left( x_n \right) \v{f}_k \left( x_n \right)^\top \v{L}^\top \sim \mathcal{W} \left( \v{V}, \Dof \right) \enspace.
\end{equation} Analogous with Eq.~\ref{eq:sigma_and_y_mvn}, this is then combined with the likelihood
\begin{equation}
     \v{y}_n \mid x_n \sim \mathcal{MVN}_\Regions\left(\v 0, \v\Sigma(x_n)\right) \enspace, \quad n = 1, \ldots, \Observations \enspace.
\end{equation}

Since the Wishart process is based on GPs, we can explicitly place model assumptions on temporal autocorrelations in the dynamic covariance. For example, if we already know that the connectivity is likely to change periodically over time, a periodic function can be used as $\kappa \left( \cdot, \cdot ;\theta \right)$. We return to the consequences of the different assumptions in Section~\ref{section:simulation_results}, and provide more details in Supplementary Section~\ref{supplementary:wishart_process}.

\subsubsection{Pooled Wishart processes} \label{subsubsection:pooled_gwp}
The Wishart process in Eq.~\eqref{eq:wishart_process} is commonly applied to model dynamic covariance based on a single multivariate time series. For example, in the case of modeling functional connectivity, the Wishart process would be applied with the fMRI time series from a single participant. However, in some scenarios, such as in the case of task-based fMRI analyses, one might be interested in an estimate of dynamic covariance at the group level, across multiple participants. The input-dependent estimate of functional connectivity $\v{\Sigma}\left( \v{x}\right)$ is then inferred from multiple multivariate time series (in our case: from multiple subjects). 

To achieve this, we extend the Wishart process by pooling observations across subjects. We expand Eq.~\eqref{eq:y_mvn} in Supplementary Section~\ref{supplementary:wishart_process} by assuming that the observations of all individual subjects share a common covariance matrix:
\begin{equation}
    \v{y}_{ns} \mid x_n \sim \mathcal{MVN}_\Regions \left( \textbf{0}, \v{\Sigma}\left(x_n \right) \right) \enspace \quad n = 1, \ldots, \Observations , \quad s = 1, \ldots, \Subjects \enspace,
\end{equation}
where $\Subjects$ refers to the number of subjects. The observations of the different subjects are assumed to be independent given the covariance. We refer to this model as the \emph{pooled Wishart process}. 

\subsection{Test statistics for dynamic connectivity} \label{subsection:test_statistics}
To determine whether a particular connection is indeed dynamic, we must first summarize the time series of connectivity using test statistics. It is important to critically assess which statistic is being used, because this formalizes which properties of the signal are considered important and encode another layer of dynamic functional connectivity assumptions~\cite{lurie2020questions}. We selected three test statistics with the aim to cover complementary aspects of connectivity dynamics. These statistics capture different properties of the dynamics, namely the overall amplitude of fluctuations in connectivity (variance)~\cite{sakouglu2010method, chang2010time}, the presence of prominent temporal frequencies~\cite{handwerker2012periodic}, and how often and strong the connectivity deviates from its median level (median-crossings)~\cite{zalesky2014time}.

\subsubsection{Variance of connectivity}
The first statistic, denoted by $\eta \left( \v{\Sigma}\left( \v{x} \right) \right)$, is the variance of the connectivity estimates. Although we compute this test statistic for each connection, we here omit the indices for notational simplicity:
\begin{equation} \label{eq:variance_test_statistic}
    \eta \left( \v{\Sigma} \left( \v{x} \right) \right) = \frac{1}{\Observations - 1} \sum_{n=1}^{\Observations} \left( \v{\Sigma}(x_n) - \mu \right)^2 \enspace.
\end{equation}
where $\v{\Sigma}(\v{x})$ is here written to refer to the connectivity between two regions over time and $\mu = \frac{1}{\Observations}\sum_{n=1}^\Observations \v\Sigma(x_n)$ is the mean of this connection. The variance increases when connections show more fluctuations over time.

\subsubsection{Maximum power across all frequencies}
The maximum power across all frequencies~\cite{handwerker2012periodic}, here written as $\psi \left( \v{\Sigma} \left( \v x \right) \right)$, is computed by first determining the power of all frequencies $\tilde{\v{\Sigma}}_{q}$ using the Discrete Fourier Transform, and then taking the largest power across all frequencies $q=1, \ldots, \Observations$:
 \begin{equation} \label{eq:maximum_power_test_statistic}
     \psi \left( \v{\Sigma} \left( \v{x} \right) \right) = \max_{q} | \tilde{\v{\Sigma}}_{q} |^2 \enspace.
 \end{equation}
The idea is that this test statistic captures periodicity in the connectivity signal. If there are periodic or near-periodic structures in connectivity over time, the power spectrum will show strong peaks at specific frequencies. Noise in the data is likely to contain low amplitudes and therefore does not produce strong peaks in the power spectrum. Hence, large powers in connectivity estimates might imply dynamics.

\subsubsection{Median-crossings}
\begin{figure}[H]
    \centering
    \includegraphics[width=\linewidth]{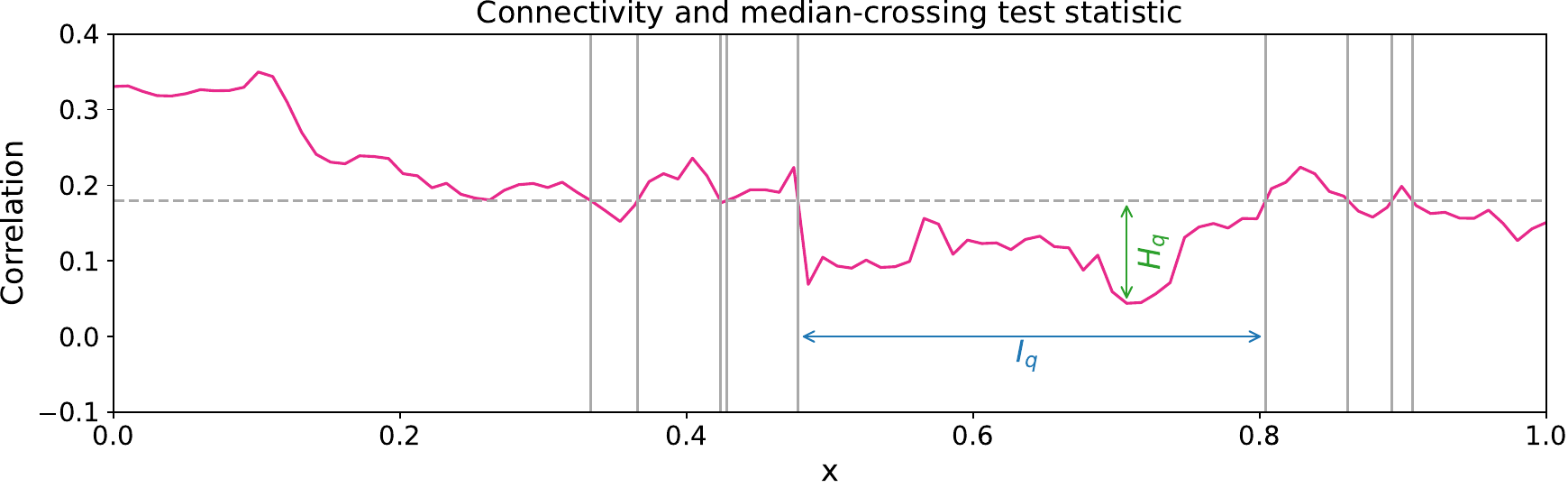}
    \caption{\textbf{An illustration of the median-crossing test statistic.} The median is shown by the dashed line, and the segments are shown by the gray vertical lines. The median-crossing test statistic is computed as a non-linear combination of the heights (in green) and lengths (in blue) of the median crossings.}
    \label{fig:median_crossing}
\end{figure}
The third test statistic evaluates the duration and magnitude of deviations from the median of the estimated connectivity~\cite{zalesky2014time} and is visualized in Figure~\ref{fig:median_crossing}. First, we obtain the median of a connection $m=\text{med}(\v\Sigma(\v x))$. Note that we here again omitted the indices to avoid clutter in the notation. Second, the connectivity estimates are divided into non-overlapping segments based on the locations where the connectivity crosses the median. Let $c_q$ be the point where the connectivity passes the median line for the $q$-th time. Then, the length of such an excursion from the median is defined as $I_{q} = c_{q+1} - c_q$, and the height as the maximum distance from the median within the window, namely $H_q = \max \left( \left| \v\Sigma(x_n) - m \right| \right)$, with $c_q \leq x_n \leq c_{q+1}$. The test statistic $\zeta \left( \v{\Sigma} \left( \v{x} \right) \right)$ is defined as a weighted sum of the heights and lengths of all segments:
\begin{equation} \label{eq:frequentist_test_2}
    \zeta\left( \v{\Sigma} \left( \v{x} \right) \right) = \sum_{q=1}^{Q-1} | I_q^\gamma H_q^\beta | \enspace,
\end{equation}
where $\gamma$ and $\beta$ are the relative weighting of the lengths and heights. This test statistic captures both the frequency of the signal (as with a higher frequency, the median is crossed more often), as well as the amplitude (as higher excursions imply a larger amplitude).

\subsection{Hypothesis testing for dynamics} \label{subsection:testing}
Here, we discuss the two different frameworks for testing for dynamic functional connectivity. The most commonly used framework for hypothesis testing is the frequentist framework, and we will compare this test against a Bayesian model comparison framework.

\subsubsection{Frequentist hypothesis testing}
\begin{figure}[H]
    \centering
    \includegraphics[width=\linewidth]{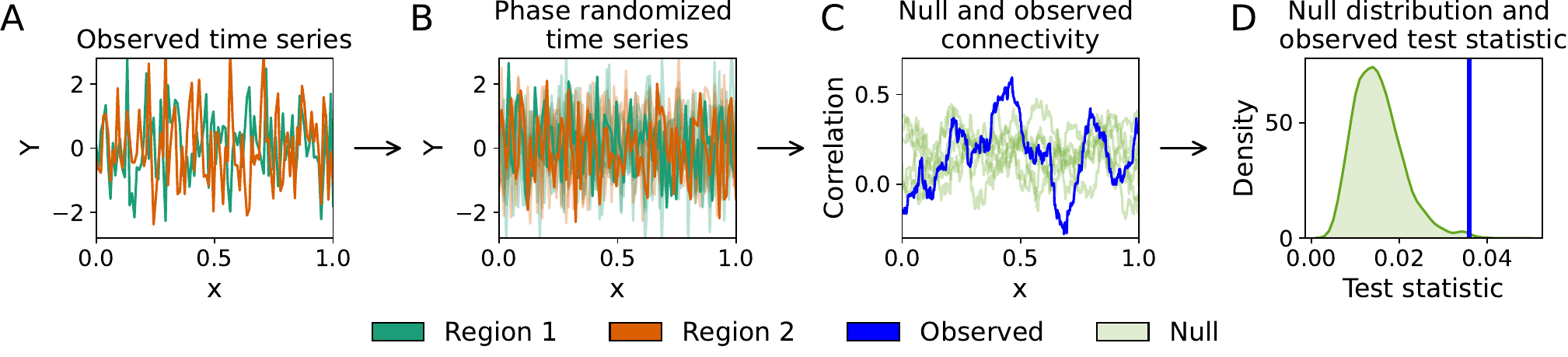}
    \caption{\textbf{An overview of the frequentist framework to statistically test for dynamic functional connectivity.} We use the observed time series (A) to generate a null distribution (B) using phase randomization. From both the observed time series and phase randomized time series, we estimate connectivity over time (C). Finally, we compute a test statistic over both, and test for significant dynamics (D).}
    \label{fig:frequentist-hypothesis-testing}
\end{figure}
Within the frequentist framework, we test for significant dynamics of the estimated dynamic connectivity by comparing it against a null distribution. The general premise is that, if the observed dynamics are unlikely given the null distribution, we conclude that the dynamics are truly present~\cite{casella2024statistical, biau2010p}. The approach consist of three main steps. First, we simulate time series data to create a null distribution and estimate its connectivity. Second, we compute a test statistic for both the actual estimate and estimates in the null distribution, and finally we compare the test statistic that corresponds to the observed signal against those of the null distribution by computing a $p$-value. This $p$-value is used to determine if the observed dynamics are significant. 

Two commonly used approaches for generating a null distribution of time series data are autoregressive randomization~\cite{liegeois2017interpreting, pervaiz2022multi} and phase randomization~\cite{hindriks2016can, handwerker2012periodic}. Both methods aim to preserve as many properties of the original time series as possible, except for the dynamics. With autoregressive randomization, an autoregressive model is estimated and subsequently applied to generate new observations. Phase randomization works as follows. First, the observed time series are transformed to the frequency domain via the discrete Fourier transform. Then, a uniformly sampled random phase $\theta\in\left[ 0, 2\pi \right]$ is added to each frequency, and finally the frequencies are transformed back into the time domain using the inverse discrete Fourier transform. Importantly, the same randomly sampled phase $\theta$ should be used for all brain regions, otherwise we also remove any static functional connectivity in between the time series. In this work, we combine the frequentist testing approach with phase randomization and the sliding-window estimation procedure. Figure~\ref{fig:frequentist-hypothesis-testing} provides a schematic overview of frequentist hypothesis testing for dynamic connectivity.

\subsubsection{Bayesian hypothesis testing} \label{subsection:bayesian_hypothesis_testing}
\begin{figure}[H]
    \centering
    \includegraphics[width=\linewidth]{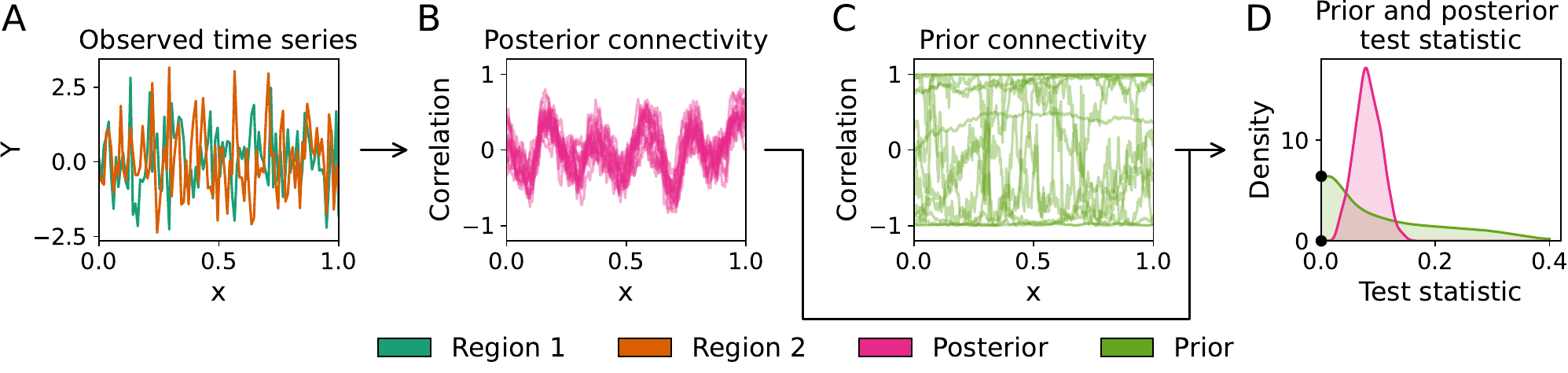}
    \caption{\textbf{An overview of the Bayesian framework to estimate functional connectivity and statistically test for dynamics.} Using the observed time series (A), the posterior distribution of connectivity is inferred (B). We compare the posterior with the prior distribution (C) to compute the Bayes factor as the fraction of the prior and posterior probability of the connectivity being static (see Eq.~\eqref{eq:savage_dickey}). These probabilities are indicated by the black dots in panel D.}
    \label{fig:bayesian-hypothesis-testing}
\end{figure}

The Bayesian estimation and testing procedure is illustrated in Figure~\ref{fig:bayesian-hypothesis-testing}. In a Bayesian estimation approach, prior beliefs about the connectivity are defined, and these beliefs are updated based on the observed time series via Bayes' theorem: 
\begin{equation}
    p \left( \v{\Sigma} \left( \v{x} \right) \mid \v{x}, \v{Y}\right) = \frac{ p\left( \v{Y} \mid \v{\Sigma} \left( \v{x} \right) \right) p\left( \v{\Sigma}\left( \v{x} \right) \right) }{p \left( \v{Y} \right)} \enspace,
\end{equation}
where $p \left( \v{\Sigma} \left( \v{x} \right) \mid \v{x}, \v{Y}\right) $ is the posterior distribution of the connectivity given the observed data, $p\left( \v{Y} \mid \v{\Sigma} \left( \v{x} \right) \right)$ the likelihood of the observed data given the connectivity, and $p \left( \v{Y} \right)$ is the marginal likelihood of the data, which serves as a normalizing constant. Importantly, $p\left( \v{\Sigma} \left( \v{x} \right) \right)$ is the prior distribution, which encodes any beliefs about the dynamic connectivity before observing any data. For example, if we expect the connectivity to smoothly vary over time, contain fast fluctuations, or have a periodic structure, we could incorporate these assumptions here. The resulting posterior estimate of connectivity is a combination of the prior assumptions and the information contained in the observed time series data. 

To test for significant dynamics in the functional connectivity, we perform Bayesian model comparison by computing Bayes factors~\cite{kass1995bayes}, which consist of the ratio of the marginal likelihoods of two models. For example, if model $\mathcal{M}_0$ assumes there is no dynamic connectivity, while model $\mathcal{M}_1$ assumes there is, then the Bayes factor is given by
\begin{equation}\label{eq:bayesfactor}
    \text{BF}_{10} = \frac{p(\v Y \mid \mathcal{M}_1)}{p(\v Y \mid \mathcal{M}_0)} \enspace,
\end{equation} indicating how much more likely the data are under $\mathcal{M}_1$ compared to $\mathcal{M}_0$. Computing the marginal likelihood of a complicated Bayesian model like the Wishart process is intractable. However, in the case of nested hypotheses, the marginal likelihood may be computed using the Savage-Dickey method~\cite{dickey1970weighted,wagenmakers_bayesian_2010}, which compares the prior and posterior probabilities at the point indicated by the null hypothesis. For instance, we can use the variance statistic from Eq.~\eqref{eq:variance_test_statistic} to test if the variance is zero (indicating no dynamics) by computing:
\begin{equation} \label{eq:savage_dickey}
    \text{BF}_{10} = \frac{p\left( \eta\left( \v{\Sigma} \left( \v{x} \right)\right) = 0 \right)}{p\left(\eta\left( \v{\Sigma}\left( \v{x} \right)\right) = 0 \mid \v{x}, \v{Y} \right)} \enspace.
\end{equation} Here, the value $\eta\left( \v{\Sigma} \left( \v{x} \right)\right) = 0$ is implicitly given by the null hypothesis: if there is no dynamic connectivity, this implies a variance of zero.

Although the Bayes factor can be thresholded to obtain a binary test outcome, it also offers valuable information about the uncertainty associated with these test outcomes. Namely, the Bayes factor quantifies the strength of evidence in favor of either the null or alternative hypothesis. To make this more clear, Eq.~\eqref{eq:savage_dickey} shows the ratio of updated beliefs after seeing the data at the null hypothesis point. If, after conditioning on the observed $\v{Y}$, our belief that $\eta\left( \v{\Sigma} \left( \v{x} \right)\right) = 0$ has increased, then we observe a Bayes factor $\text{BF}_{10}<1$, indicating evidence in favor of the null hypothesis. On the other hand, if our beliefs about the possibility $\eta\left( \v{\Sigma} \left( \v{x} \right)\right) = 0$ have decreased, then the data suggest the null hypothesis has become less likely, and the alternative model has gained evidence instead, with $\text{BF}_{10}>1$. According to general practices~\cite{kass1995bayes, hoijtink2019tutorial, huth2025statistical}, log Bayes factors greater than 3 are considered conclusive evidence for the alternative hypothesis, whereas values below 1/3 indicate strong evidence for the null hypothesis. Based on these guidelines, we can classify a connection as dynamic if its log Bayes factor is above 3.

\subsection{Simulations}
Since a ground truth is usually unknown in empirical time series, we first compare the reliability of the frequentist sliding-window and Bayesian Wishart process hypothesis testing approaches on simulated time series with different characteristics. Each simulation consists of $\Regions=2$ brain regions and $\Observations \in \{150,300, 600\}$ observations. We let the time range $\v{x}$ be uniformly spaced from $0$ to $1$. To simulate time series data, we first generate correlation matrices of which the off-diagonal element is either static or dynamic, and the diagonal elements are set to 1: 
\begin{equation} \label{eq:sigma_simulation}
    \v{\Sigma}(x_n) = \begin{pmatrix} 1 & \sigma(x_n) \\ \sigma(x_n) & 1\end{pmatrix} \enspace.
\end{equation}
Next, the time series were generated by sampling from a multivariate distribution using:
\begin{equation} \label{eq:y_mvn}
    \v{y}_n \mid x_n \sim \mathcal{MVN}_\Regions \left( \v{0}, \v{\Sigma}\left(x_n\right) \right) \enspace, \quad n = 1, \ldots, \Observations \enspace.
\end{equation}

Static connections were set to a constant value within the range $[-0.4, 0.4]$. For the dynamic simulations, we simulated three different types of connectivity patterns, namely periodic, state-switching and fMRI-like structures. In the periodic simulation, the off-diagonal element of $\v{\Sigma}(x_n)$ follows a sine wave with varying amplitudes $A \in \{ 0.2, 0.4, 0.6, 0.8 \}$ and frequencies $\omega \in \{1, 2, \ldots, 5\}$:
\begin{equation}
    \sigma(x_n) = A \sin \left(\omega 2\pi x_n + \varphi \right) \enspace, \quad n = 1, \ldots \Observations \enspace.
\end{equation}
The phase $\varphi$ was randomly set to a value within the range $[0, 2\pi]$.

The state-switching simulation was based on the work by Thompson et al.~\cite{thompson2018simulations}. Namely, we let the covariance matrix in Eq.~\eqref{eq:sigma_simulation} switch between two states with the strength of connectivity $\sigma(x_n)$ being $0.1$ in one state and $0.8$ in the other state. The duration of each state is randomly sampled from $\left[20, 30, 40, 50, 60 \right]$ time points. After the sampled number of time points, the simulation switches to the other state. 

Finally, the fMRI-like simulation was based on the work by Kampman et al.~\cite{kampman2024time}. Here, we first convolve the state-switching pattern with a haemodynamic response function (HRF), and then add realistic noise based on fMRI timeseries data from the Human Connectome Project (HCP) dataset~\cite{van2013wu}. Namely, in these simulations, noise (denoted by $\v{\epsilon}_n$) is added to the time series as follows:
\begin{equation} \label{eq:y_fmri}
    \v{y}_n^* = \alpha \v{y}_n + (1 - \alpha) \v{\epsilon}_n \enspace, \quad n = 1, \ldots, \Observations \enspace,
\end{equation}
where we set $\alpha$ to $2/3$ such that the signal to noise ratio equals 2. Similarly to the work by Kampman et al.~\cite{kampman2024time}, we randomly selected independent components from an independent component analysis (ICA) on the HCP resting-state fMRI dataset. In this way, we can be sure that only the signal to noise ratio in the data is affected, but not the pattern of the covariance itself. The variance and covariance terms of the new timeseries are only scaled by the signal to noise ratio~\cite{kampman2024time}.

We generate 40 random connectivity matrices over time for each combination of simulation parameters, that is for every connectivity pattern and number of observations, and in the case of periodic connectivity also for every amplitude and frequency. Using Eq.\eqref{eq:y_mvn}, we then generate one random dataset for each of these simulated connectivity matrices over time. Implementations of these simulations and the corresponding hypothesis tests can be found on \href{https://github.com/Hesterhuijsdens/Bayesian-dynamic-connectivity-hypothesis-testing.git}{our GitHub page}. 

\subsection{Participants, fMRI data, and parcellation} \label{subsection:fmridata}
To explore how the two frameworks can be used in practice to test for dynamics, we test for dynamics in a working memory task dataset from the Human Connectome Project (HCP)~\cite{van2013wu}. For computational reasons, we used the 100 unrelated subjects subset of the HCP database. From these 100 subjects, we only selected participants with a task accuracy above 60\%, resulting in fMRI data from 95 healthy participants. 

The task paradigm is described in Supplementary Section~\ref{supplementary:working_memory_task}. In the n-back task, the 0-back blocks are considered to require a lower memory load than the 2-back blocks. Previous studies have found that increase in working memory load is associated with an increase in functional connectivity within the frontoparietal network, and a decreased connectivity in the default mode network~\cite{piccoli2015default}.

The fMRI data was parcellated using Glasser's MMP atlas~\cite{glasser2016multi}. To select regions of interest, we used the sliding-window approach with a window size of 10\% of the number of observations to compute the correlation between each region pair's functional connectivity and the 2-back task paradigm. We then selected the pair of regions with the strongest correlation, namely the dorsolateral prefrontal cortex (DLPFC) and the inferior parietal lobule (IPL). Additionally, we selected another region that was unrelated to the task paradigm, namely the primary auditory cortex (A1). We expect that especially the connectivity between the DLPFC and IPL regions is involved in working memory and hence changes dynamically depending on the task condition~\cite{barch2013function, sambataro2010age}. The connectivity between the DLPFC and A1 regions and between the IPL and A1 regions are expected to change less throughout the task paradigm. 

\subsection{Assumptions on connectivity}
In this work, the frequentist testing approach is combined with the sliding-window method and the Bayesian approach with the Wishart process from Eq.~\eqref{eq:wishart_process}. That is, we first estimate functional connectivity using these methods, and then apply a hypothesis test to the resulting connectivity estimates. Both the sliding-window method and Wishart process require the user to define certain modeling choices. 

\subsubsection{Sliding-window assumptions}
In the sliding-window method, we set the window length to a fraction of the number of observations to ensure the window captures a similar part of the latent correlations. We compare three different lengths, namely $5\%$, $10\%$ and $20\%$ of the number of observations in the simulation. In the HCP working memory dataset, this comes down to 20, 40 and 80 time points, which correspond to 14.4, 28.8 and 57.6 seconds of recording. The stride length $\tau$ was set to $1$ TR (0.72 seconds), giving us an estimate of functional connectivity at every time point. By setting the stride length to $1$, windows will overlap substantially, and therefore we encode the assumption that the functional connectivity varies smoothly over time. Additionally, short window sizes assume that the connectivity will change rapidly, at the expense of signal-to-noise in each window and hence more noisy estimates. Longer window sizes will result in more stable estimates of connectivity, but might miss meaningful fluctuations. Finally, for the frequentist testing framework, we generated 1000 surrogate datasets to construct the null distribution. For the median-crossing statistic, we set the respective weights of the excursion lengths and heights to $\gamma=0.9$ and $\beta=1$, following previous work~\cite{hindriks2016can,zalesky2014time}.

\subsubsection{Wishart process assumptions}
For the Wishart process, a GP kernel, and priors on the kernel hyperparameters $\theta$ and the scale matrix $\v{V}$ need to be defined. 

\begin{figure}[H]
    \centering
    \includegraphics[width=\linewidth]{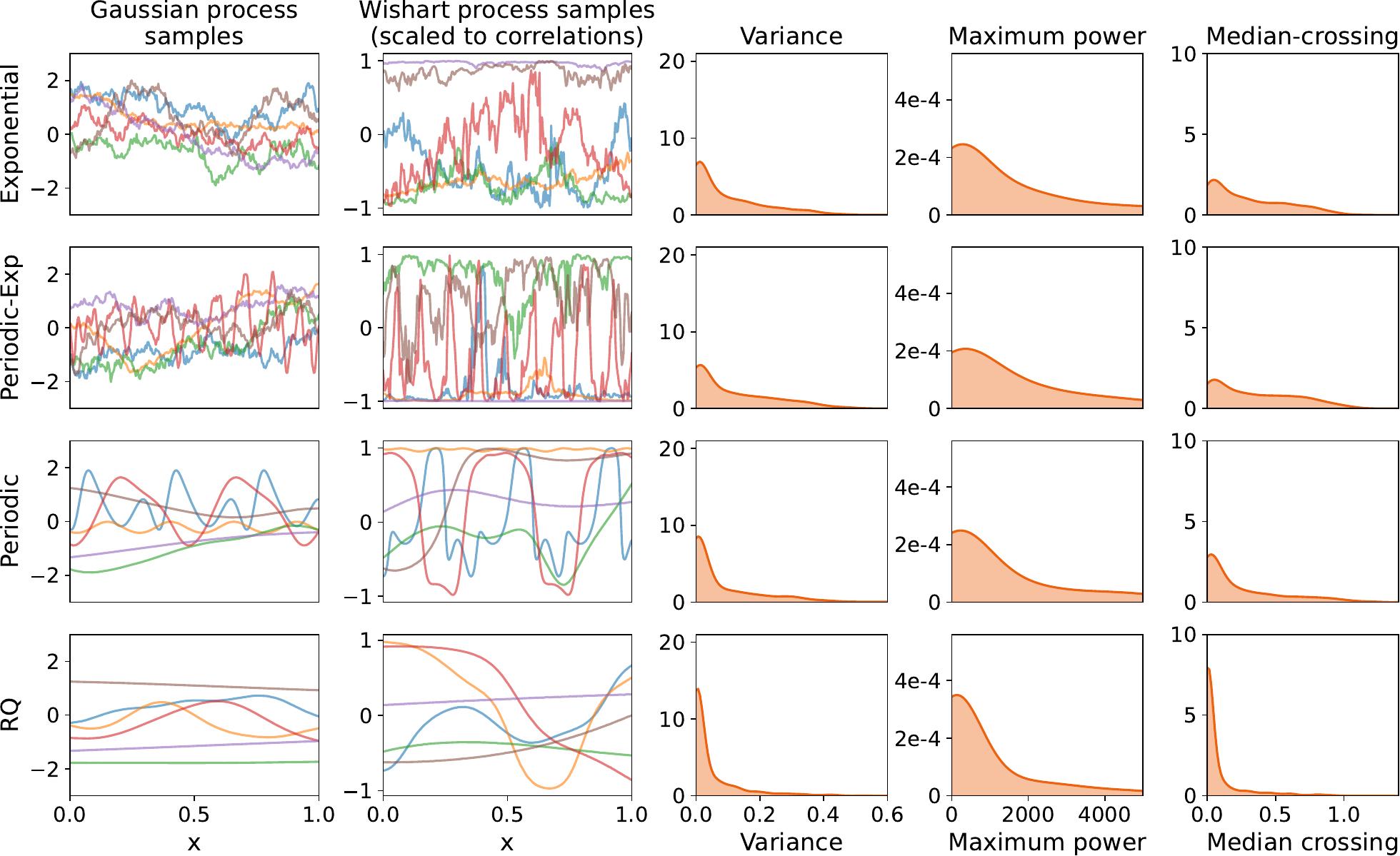}
    \caption{\textbf{Samples from the Wishart process prior and their corresponding prior Gaussian process samples and }test statistic distributions. Four different kernel and parameter initializations are shown. All kernel hyperparameter priors were set to a log-normal with mean of 0 and standard deviation of 1. The prior on the individual elements of the lower Cholesky decomposition of the scale matrix was set to be a normal distribution with mean of 0 and standard deviation of 1. Although the Gaussian process and Wishart process samples share similar properties, the link between these two is not straightforward.}
    \label{fig:null-distributions-for-different-kernels}
\end{figure}
To explore the effect that different kernels can have, we use four different GP kernels in our experiments, namely an exponential, periodic-exponential (often referred to as a locally periodic), periodic, and rational quadratic kernel:
\begin{equation} \label{eq:kernels}
    \begin{split}
        &\kappa_\text{exponential} \left( x_n, x_m ; \ell_\text{exp} \right) = \exp \left( - \frac{\left| x_n - x_m \right| }{2\ell_\text{exp}^2} \right) \enspace, \\
        &\kappa_\text{periodic-exponential} \left( x_n, x_m ; p, \ell_\text{periodic}, \ell_\text{exp} \right) = \exp \left( - \frac{2\sin^2 \left( \pi \left| x_n - x_m \right| / p \right) }{\ell_\text{periodic}^2} \right) \exp \left( - \frac{\left| x_n - x_m \right| }{2\ell_\text{exp}^2} \right) \enspace, \\
        &\kappa_\text{periodic} \left( x_n, x_m ; p, \ell_\text{periodic} \right) = \exp \left( - \frac{2\sin^2 \left( \pi \left| x_n - x_m \right| / p \right) }{\ell_\text{periodic}^2} \right) \enspace, \\
        &\kappa_\text{RQ} \left( x_n, x_m ; \alpha, \ell_\text{exp} \right) = \left( 1 + \frac{\left| x_n - x_m \right|}{2\alpha \ell_\text{exp}^2} \right)^{-\alpha} \enspace.
    \end{split}
\end{equation}
The periodic-exponential kernel is formed by multiplying an exponential and a periodic function. The different GP covariance functions and kernel parameters place assumptions on the properties of the latent GP samples, and therefore also on the characteristics of the functional connectivity being estimated. For example, a periodic kernel assumes a repeating pattern in the connectivity structure over time, whereas an exponential kernel assumes functional connectivity with non-smooth fluctuations over time. These assumptions are shown in more detail in Figure~\ref{fig:null-distributions-for-different-kernels}, where samples from the prior of the Gaussian process and Wishart process are shown for different kernels and kernel hyperparameters, together with their corresponding a priori test statistic distributions. As illustrated in the figure, the Gaussian process and Wishart process samples share similar properties based on the kernel and hyperparameters, but the exact link between the two is not straightforward. Connectivity samples from the exponential and periodic-exponential kernels contain smaller fluctuations and are less smooth than the periodic and rational quadratic kernels. This is also influenced by the kernel hyperparameters. The exponential, periodic-exponential and rational quadratic kernels have a lengthscale parameter $\ell_\text{exp}$, and both the periodic and periodic-exponential kernels have parameters $p$ and $\ell_\text{periodic}$, which represent the period and lengthscale within a period. The rational quadratic kernel has an additional $\alpha$ parameter that determines the weighting of small and large variations. If $\alpha$ increases, GP samples of this kernel will become increasingly smooth. In the prior, all hyperparameters were assumed to follow a log-normal prior with mean of 0 and standard deviation of 1. We set a normal prior with mean of 0 and standard deviation of 1 on each element of the lower Cholesky decomposition ($\v{L}$) of the scale matrix $\v{V}$. Additionally, we explore the effects of alternative priors on the hyperparameters and scale matrix on the performance in Supplementary Section~\ref{supplementary:robustness}.

\section{Results of simulation studies} \label{section:simulation_results}
First, we show a few representative individual connectivity estimates by the sliding-window method and Wishart process, and we discuss the ability of both methods to detect dynamic functional connectivity. We compare strengths and limitations of the frequentist and Bayesian approaches in different simulations. Then, we explore how uncertainty quantification, provided by the Bayesian framework, can provide additional insights into dynamics. In this section, we focus on the periodic simulations. All results for the other simulations and test statistics can be found in Supplementary Section~\ref{supplementary:additional_simulations}. 

\subsection{Connectivity estimates} \label{subsection:connectivity_estimates}
\begin{figure}[H]
    \centering
    \includegraphics[width=\linewidth]{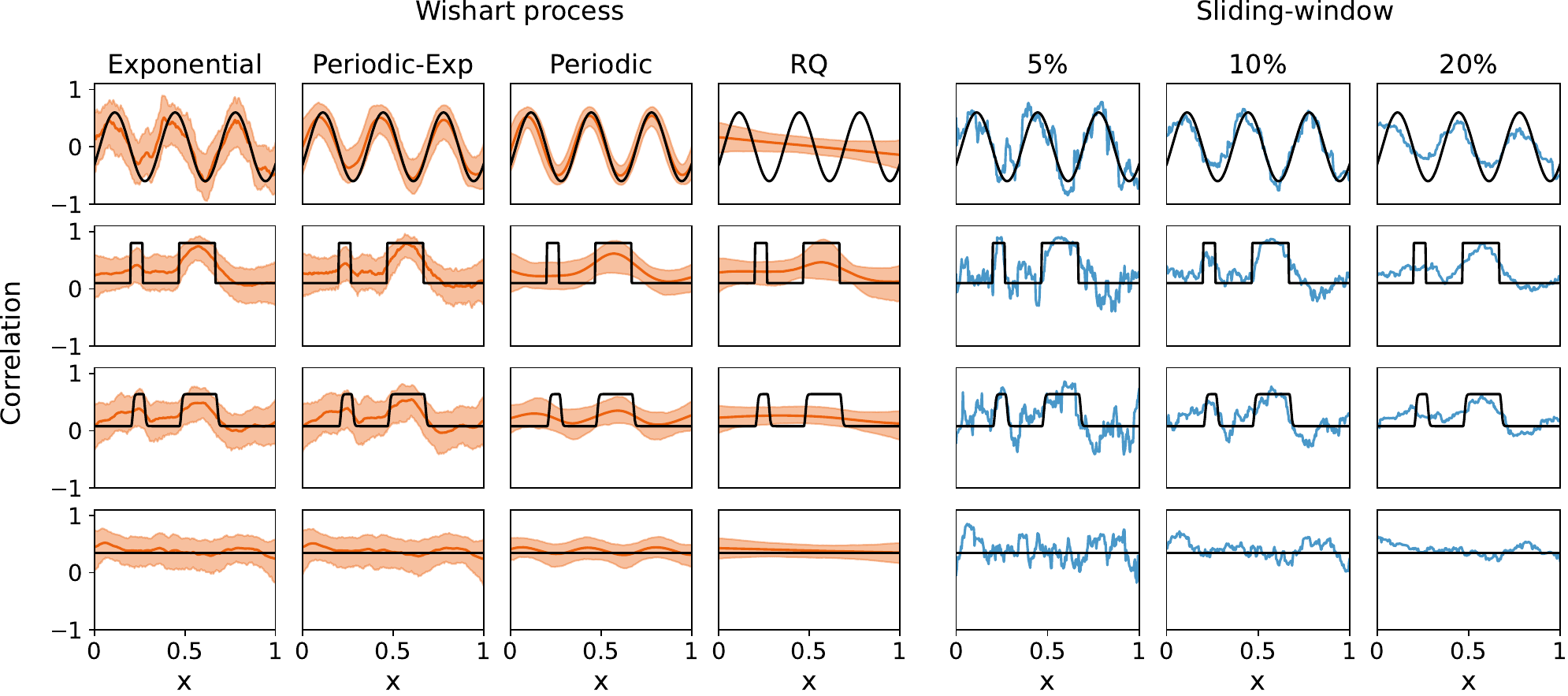}
    \caption{\textbf{Examples of connectivity estimates based on simulated data with $\Observations=300$ observations.} We show estimates based on all four types of simulations, namely periodic connectivity with a frequency of 3 and an amplitude of 0.6, state-switching, fMRI-like, and static simulations. We show the estimates using different kernels and window lengths. For the Wishart process, we show the 95\% highest density interval. The ground truth correlations are shown in black.}
    \label{fig:connectivity_estimates}
\end{figure}
Figure~\ref{fig:connectivity_estimates} presents a few examples of connectivity estimates by the Wishart process and sliding-window method using different kernels and window lengths. These results indicate that both methods can accurately capture the latent connectivity, but that the quality of these estimates is affected by the choice of window length and kernel. For example, when using a rational quadratic kernel, the Wishart process is unable to capture the fast fluctuations in the connectivity. Estimates become more accurate when assuming periodicity, or when using the exponential kernel. The sliding-window estimate, with a window length of 20\% of the observations, seems too smooth to capture the rapid fluctuations in the connectivity. 

The second row illustrates a latent connectivity with non-periodic state switches. In this case, the sliding-window method captures the rapid switches in connectivity more accurately than the Wishart process. Nonetheless, the performance differs based on the three window lengths, as a window length of 20\% is too smooth to model the rapid switches, whereas a window length of 5\% estimates a lot of fluctuations in the static periods of the connectivity. 

The third row shows the connectivity estimates based on the simulated fMRI timeseries. Overall, we observe similar estimates as for the state-switching simulation. However, there seems to be a slight decrease in accuracy, especially for the Wishart process with a rational quadratic kernel.

Finally, the bottom row presents the static connectivity scenario. Small window lengths estimate a lot of fluctuations, whereas larger window lengths provide more accurate estimates of connectivity. The Wishart process captures the static connection well, as the zero line is fully encapsulated by the 95\% highest density areas.

The accuracy of the connectivity estimates across all simulations is provided in Supplementary Section~\ref{supplementary:all_mses} and is consistent with the example in Figure~\ref{fig:connectivity_estimates}. Both the window size and kernel affect the quality of the connectivity estimates, where large window sizes and smooth kernels are unable to capture fast fluctuations in connectivity. In the state-switching and fMRI-like simulations, the Wishart process estimates with a periodic or periodic-exponential kernel outperform those from the sliding-window method. The same is true for static connections, which are challenging for the sliding-window approach.

\subsection{Hypothesis test performances} \label{subsection:performance}
To compare the frequentist and Bayesian hypothesis testing frameworks in detecting dynamic connectivity, we first convert the Bayes factors obtained from Eq.~\eqref{eq:savage_dickey} into binary decisions. Based on general guidelines mentioned in Section~\ref{subsection:bayesian_hypothesis_testing}, we classify a connection as dynamic if its log Bayes factor is above 3. We then evaluate both frameworks based on three different metrics, namely accuracy, recall, and false positive rate. Accuracy provides a measure of the overall number of correctly classified connections, taking into account both static and dynamic connections. Recall quantifies the number of connections that is correctly classified as being dynamic, so it measures the sensitivity of the hypothesis test. The false positive rate indicates the number of connections incorrectly classified as being dynamic. In the frequentist framework, the false positive rate is directly related to the significance threshold. Since we classify a p-value below 0.05 as significantly dynamic, the false positive rate of this framework will approximate 5\%. 

\begin{figure}[H]
    \centering
    \includegraphics[width=\linewidth]{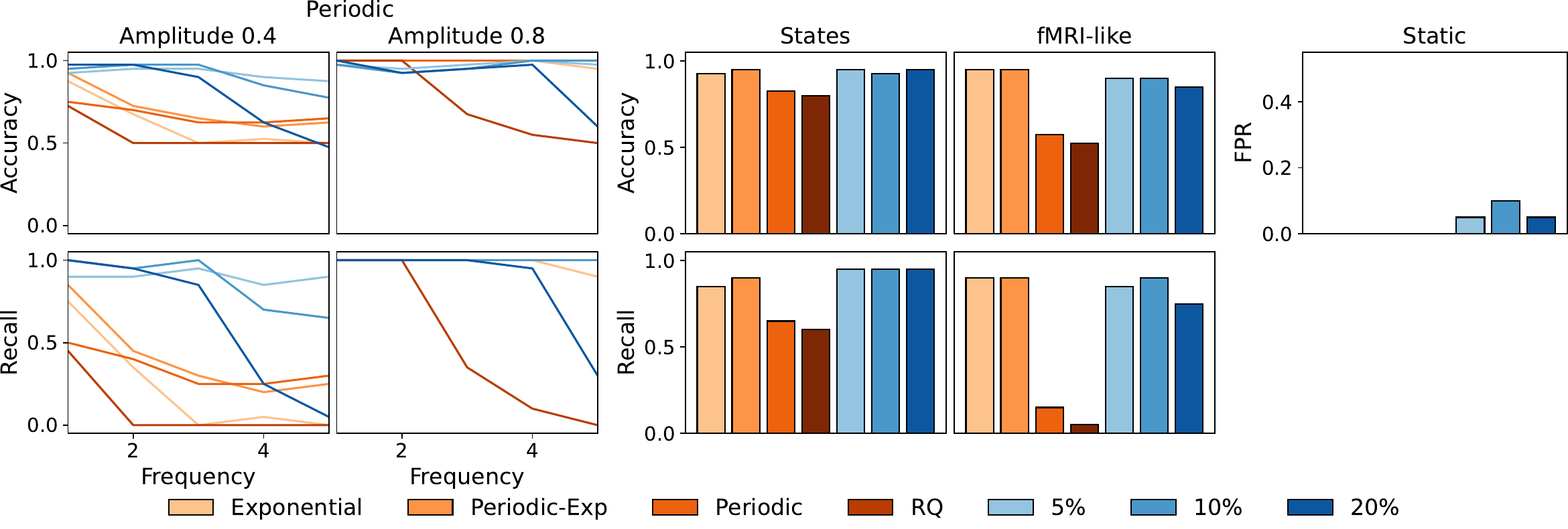}
    \caption{\textbf{Hypothesis test performances based on the variance test statistic.} The figure shows the fractions of correctly classified connections out of all connections (accuracy) and out of only dynamic connections (recall), and the fraction of connections classified incorrectly as dynamic (false positive rate, FPR), when using the variance test statistic.}
    \label{fig:results-hypothesis-testing}
\end{figure}
Figure~\ref{fig:results-hypothesis-testing} shows the accuracy, recall and false positive rate for the simulations with $\Observations=300$ observations based on the variance test statistic. We show the state-switching, fMRI-like, static and periodic simulations. For the periodic simulations, only the results for amplitudes of $0.4$ and $0.8$ are shown, but all results for other amplitudes, test statistics and numbers of observations are shown in Supplementary Section~\ref{supplementary:additional_simulations}. 

As expected, recall improves with more observations and higher amplitudes. Generally, the frequentist approach tends to outperform the Bayesian approaches in these simulations, particularly for recall and to a lesser degree in accuracy. However, modeling choices also have a large effect on performance. In the periodic simulations, the periodic and periodic-exponential kernels, as well as smaller window sizes of 5\% and 10\%, perform well, even for faster dynamics. However, the rational quadratic and exponential kernels, and the largest window size (20\%), perform well on low-frequency dynamics, but not always on higher frequencies. In the state-switching simulations, the exponential and periodic-exponential kernels tend to perform best, along with the different sliding-window approaches. A similar pattern can be observed for the fMRI-like simulations. However, compared to the state-switching simulation, the recall of the fMRI-like simulation has decreased for both the sliding-window and the Wishart process.

\subsection{Uncertainty in Bayesian hypothesis testing} \label{subsection:uncertainty}
\begin{figure}[H]
    \centering
    \includegraphics[width=\linewidth]{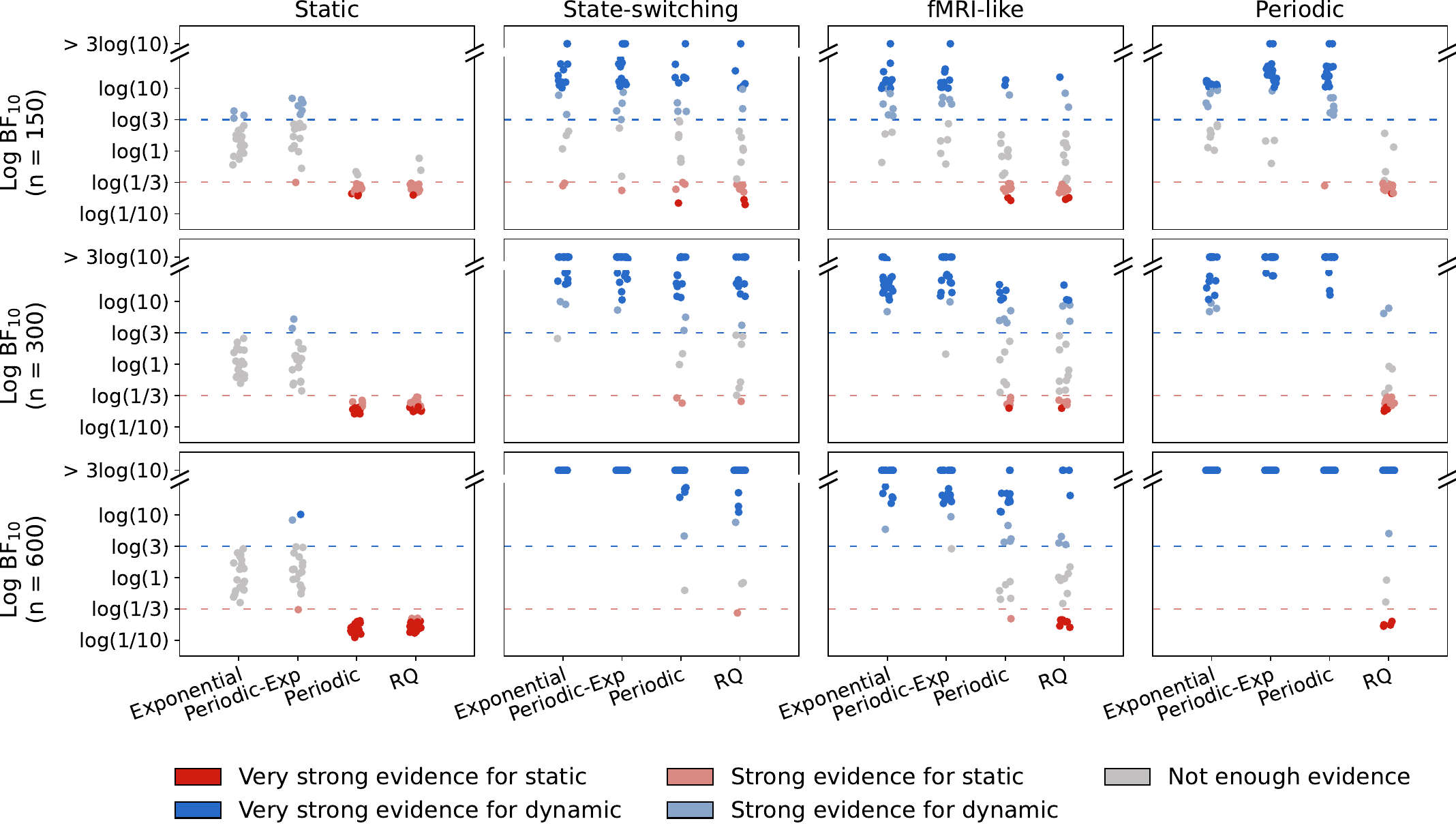}
    \caption{\textbf{Log Bayes factors of detecting dynamics based on the variance test statistic.} The figure shows the results for the static, state-switching and fMRI-like simulations, and the periodic simulations with a frequency of 3 and an amplitude of 0.6. Every dot represents a single connection and is colored based on the amount of evidence for the connection being dynamic or static. Thresholds for significant evidence are indicated by the dashed lines.}
    \label{fig:bayes-factors}
\end{figure}
Figure~\ref{fig:bayes-factors} presents the distributions of log Bayes factors for the static, state-switching and fMRI-like simulations and for the periodic simulations with a frequency of 3 and an amplitude of 0.6. Results for the other periodic simulations and test statistics can be found in Supplementary Section~\ref{supplementary:additional_simulations}. Each point in the figure represents a single connection and is colored based on the strength of its evidence.

For all simulations, the Bayes factors increase with the number of observations. For $\Observations=150$ observations, a large number of edges is inconclusive. Moreover, the results of the static simulations show that, depending on the kernel choice, the strength of evidence differs. Even with $\Observations=600$ observations, the strength of evidence for most edges estimates by the exponential and periodic-exponential kernels remains inconclusive. The results on the state-switching and fMRI-like simulations further support this finding that a sufficient number of observations is needed to find strong evidence for dynamics. In the results of the periodic simulations, both in Figure~\ref{fig:bayes-factors} and in Supplementary Section ~\ref{supplementary:additional_simulations}, it can be seen that the number of connections with conclusive evidence increases with an increasing amplitude and a decreasing frequency. This is in line with the recall scores from Figure~\ref{fig:results-hypothesis-testing}. Connections that are estimated less accurately, such as the periodic simulations with frequency 3 using the rational quadratic kernel (see also Supplementary Figure~\ref{supplementary:mse_n300_mean_estimates}), result in inconclusive evidence as well.

Overall, these results indicate that both the amount of available observations and the modeling choices strongly influence the outcomes of the hypothesis tests. With a limited number or observations or a less suitable modeling choice, such as a smooth kernel while the connectivity has fast fluctuations, it may not be possible to find enough evidence for detecting a dynamic functional connection.

\section{Empirical results} \label{section:empirical_results}
We applied both statistical frameworks to the n-back working memory task fMRI dataset from the Human Connectome Project~\cite{van2013wu} to illustrate the behavior of hypothesis testing for dynamics in an empirical setting. As discussed in Section~\ref{subsection:fmridata}, prior work suggests a difference in functional connectivity between high and low working memory load within the frontoparietal network and the default mode network, as these regions are found to be consistently associated with working memory~\cite{barch2013function, piccoli2015default}. Specifically, we explore whether functional connectivity between the dorsolateral prefrontal cortex (DLPFC) and the inferior parietal lobe (IPL) increases during high working memory load (in the 2-back task), as compared to low working memory load (in the 0-back task). Importantly, we explore whether this connectivity changes dynamically over the task paradigm. For comparison, we also explore the connections between the primary auditory cortex (A1) and the IPL, and between the DLPFC and the A1. While we expect less task-related dynamics between the primary auditory cortex (A1) and the IPL and between the DLPFC and the A1, some dynamics might still be present due to  increases in IPL and DLPFC activity that are driven by the task. Importantly, because the true connectivity patterns are unknown, the goal here is not to compare the accuracy of the different methods. Instead, this analysis is intended to explore how the two frameworks perform in an empirical setting.

\subsection{Connectivity estimates}
\begin{figure}[H]
    \centering
    \includegraphics[width=\linewidth]{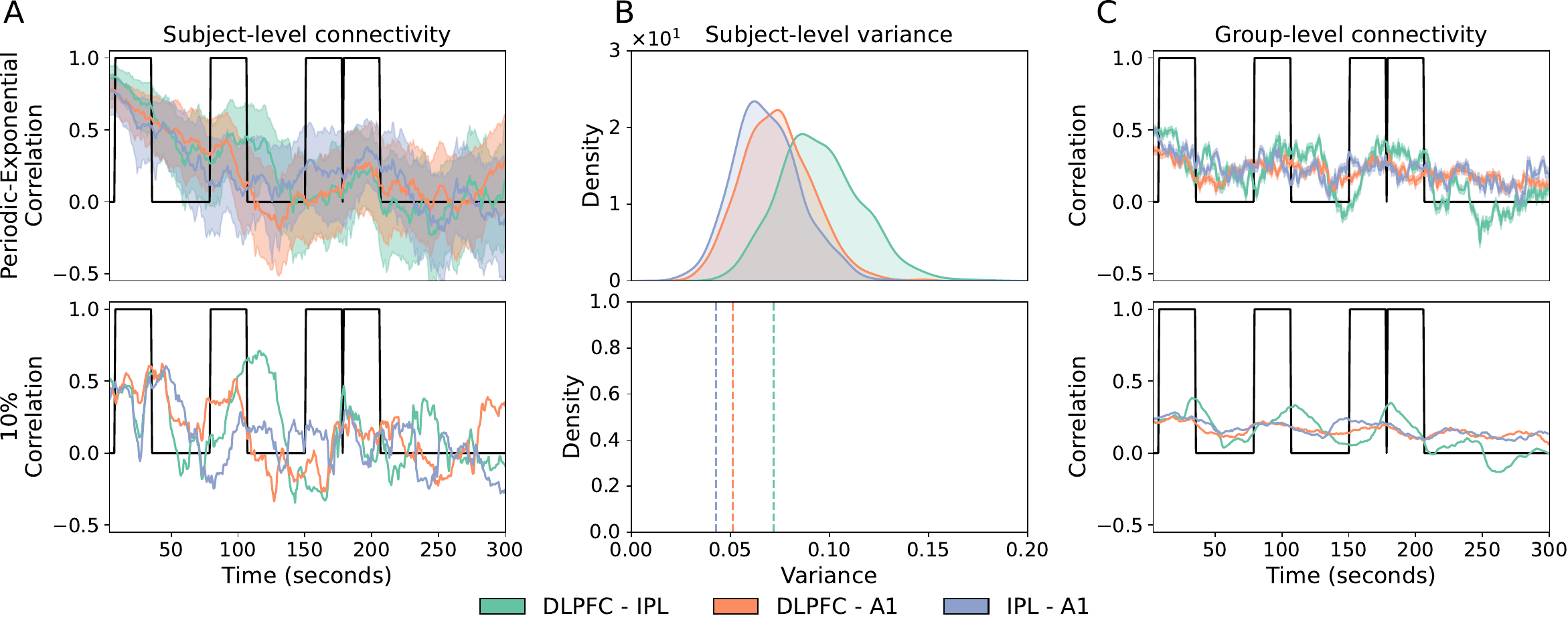}
    \caption{\textbf{Estimates of functional connectivity between the dorsolateral prefrontal cortex (DLPFC), inferior parietal lobe (IPL) and primary auditory cortex (A1), and its corresponding variance.} Subplot A shows the connectivity estimates based on a single representative subject, and subplot B shows the corresponding variance test statistic. For the Wishart process, we have a full posterior distribution over the test statistic, whereas for the sliding-window we have a point estimate. Subplot C shows the estimates on the group level (the corresponding group-level statistics are shown in Fig.~\ref{fig:test_statistics_empirical}). The 2-back task design is shown in black.}
\label{fig:connectivity_estimates_and_test_statistics_empirical}
\end{figure}
Figure~\ref{fig:connectivity_estimates_and_test_statistics_empirical} visualizes a few examples of estimates of functional connectivity between the DLPFC, IPL and A1 regions. Figure~\ref{fig:connectivity_estimates_and_test_statistics_empirical}A shows these estimates based on an individual representative subject for both the Periodic-Exponential kernel and a window size of 10\%. To quantify how the subject-level functional connectivity estimates fluctuate with the HRF-convolved task paradigm, we computed their correlation. For the Wishart process estimates, we used the mean of the posterior estimates. Averaged over all subjects, the correlations for the Wishart process were $r=0.246\pm0.200$ for the DLPFC--IPL connection, $r=0.119\pm0.219$ for the DLPFC--A1 connection, and $r=0.115\pm0.221$ for the IPL--A1 connection. For the sliding-window, these were $r=0.227\pm0.215$, $r=0.106\pm0.237$, and $r=0.107\pm0.216$, respectively. Figure~\ref{fig:connectivity_estimates_and_test_statistics_empirical}B shows the corresponding posterior distributions or point estimates of the variance, indicating that the differences between regions are indeed quite small for the Wishart process.

Additionally, we estimated functional connectivity based on the group level, using the pooled Wishart process from Section~\ref{subsubsection:pooled_gwp}. For the sliding-window method, the group-level estimates were obtained by averaging across all individual subject estimates. A few representative examples of estimates are shown in Figure~\ref{fig:connectivity_estimates_and_test_statistics_empirical}C. At the group level, the Wishart process showed correlations of $r=0.558$ for the DLPFC--IPL connection, $r=0.447$ for DLPFC--A1, and $r=0.403$ for IPL--A1. Similarly, the sliding-window method showed correlations of $r=0.568$, $r=0.524$, and $r=0.458$ for the DLPFC--IPL, DLPFC--A1 and IPL--A1 connections. Overall, this suggests that the fluctuations with the task paradigm are not uniquely present in the DLPFC--IPL connection, but they do correlate more strongly. Moreover, compared to the estimates on a single subject, the uncertainty in the connectivity estimates is lower at the group level.

\subsection{Differences in amount of dynamics in functional connectivity}
\begin{figure}[H]
    \centering
    \includegraphics[width=\linewidth]{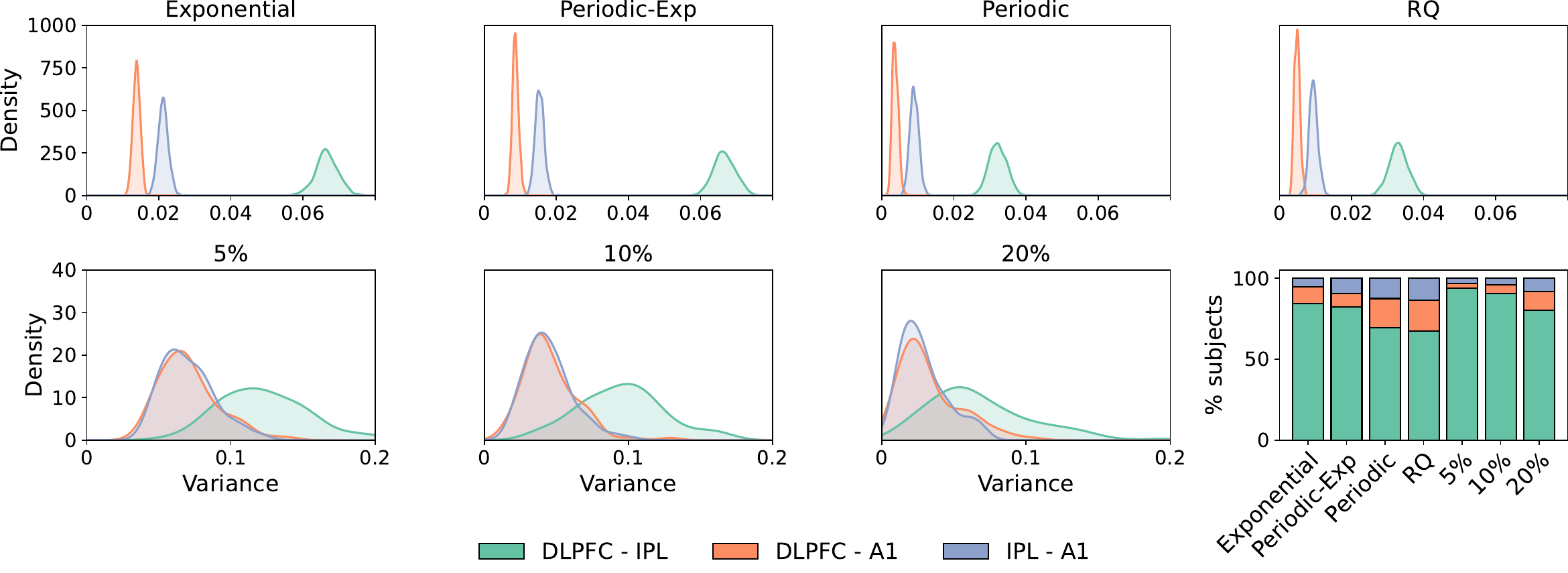}
    \caption{\textbf{Subject-level ordering of variances and group-level variances for three pairs of brain regions from the working memory task.} The barplot indicates the number of times (in percentage of all subjects) that each pair of regions has the largest value, indicating that this connection is most dynamic. The density plots show the variance by means of the posterior distributions on the group level (by the pooled Wishart process model) and the distributions of the sliding-window estimates across all subjects.}
    \label{fig:test_statistics_empirical}
\end{figure}
In Figure~\ref{fig:test_statistics_empirical}, the variance test statistic is shown for all three connections under different modeling choices. The barplot in Figure~\ref{fig:test_statistics_empirical} summarizes the subject-level results by the percentage of subjects in which each connection showed the largest variance. For the Wishart process, we used the posterior mean. In line with our expectations, the DLPFC--IPL connection was most often ranked as the most dynamic connection, but these results varied across kernels and window sizes. 

Additionally, the density plots in Figure~\ref{fig:test_statistics_empirical} show the group-level results across all subjects. For the Wishart process, these results are based on the posterior distributions inferred by the pooled Wishart process from Section~\ref{subsubsection:pooled_gwp}. For the sliding-window method, the group-level distributions were obtained by applying kernel density estimation over the individual subject estimates. In general, these results are in line with our prior expectations. Namely, the variance of the DLPFC--IPL connection is consistently higher than the DLPFC--A1 and IPL--A1 connections across all kernels and window sizes. This suggests that the DLPFC--IPL connection is indeed most dynamic.

Results based on the median-crossing and maximum power test statistics can be found in Supplementary Section~\ref{supplementary:additional_empirical}. The results based on the maximum power are in line with our findings based on the variance, showing a clear preference for the DLPFC--IPL connection. A similar pattern can be observed for the median-crossings, but the differences between connections are smaller. Overall, the empirical results show that the DLPFC--IPL connection is consistently found to be most dynamic at the group level, but results on the individual level vary largely across modeling choices, and no ground truth regarding the true dynamics is known. The results highlight the behavior of different prior modeling choices in an empirical setting.

\begin{figure}[H]
    \centering
    \includegraphics[width=\linewidth]{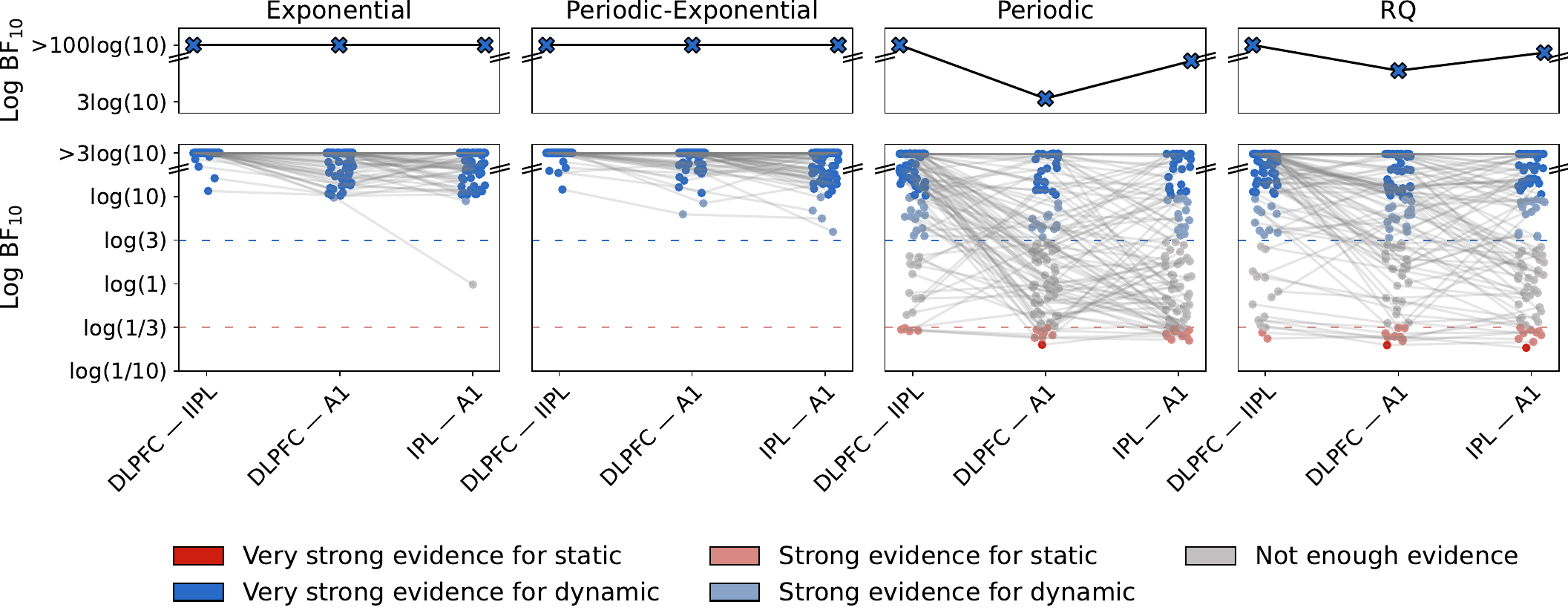}
    \caption{\textbf{Log Bayes factors of detecting dynamics in functional connectivity during the n-back task.} The crosses in the top row indicate the evidence in the connectivity across all subjects, as estimated by the pooled Wishart process. Every dot represents a connection of a single subject and is colored based on the amount of evidence of the connection being either dynamic or static. The lines between these dots indicate the corresponding pairs of brain regions of each subject. Thresholds for significant evidence are indicated by the dashed lines.}
    \label{fig:bayes_factors_empirical}
\end{figure}
While in Figures~\ref{fig:connectivity_estimates_and_test_statistics_empirical} and ~\ref{fig:test_statistics_empirical} we showed that the DLPFC--IPL connection shows a higher variance test statistic than the other two connections, this does not yet show that this connection is truly dynamic. In Figure~\ref{fig:bayes_factors_empirical}, we present the log Bayes factors for all kernels and connections based on the variance test statistic. The crosses in the top rows indicate the group-level Bayes factors, whereas Bayes factors of individual participants are indicated by dots in the bottom rows, with gray lines connecting the values across region pairs. In general, the DLPFC--IPL connection shows stronger evidence for dynamics than the other two connections. However, on the individual level there are large differences across modeling choices. This becomes especially clear from the wide spread in Bayes factors when using the periodic kernel. 

Overall, these results highlight how different modeling choices can largely influence the conclusions about functional connectivity being dynamic. At the group level, the DLPFC--IPL connection clearly shows stronger evidence for dynamics than the other two connections. Here, results are more consistent across modeling choices.

\section{Discussion} \label{section:discussion}
There is an increasing interest in modeling dynamic functional brain connectivity, in which the strength of connectivity does not remain static, but varies as a function of some input variable, such as time~\cite{kampman2024time, liegeois2019resting, lurie2020questions, ahrends2022data}. To determine if this dynamic connectivity reflects true dynamics, or only fluctuates due to noise or modeling choices, statistical testing is needed~\cite{laumann2017stability, hindriks2016can}. In this work, we introduced a Bayesian hypothesis testing framework for dynamic connectivity, which makes use of Wishart processes to model functional connectivity over time~\cite{kampman2024time, heaukulani2019scalable, wilson2010generalised, huijsdens2024robust}. Unlike in the frequentist alternative combined with the sliding-window method, which is commonly used to test for dynamics and where modeling assumptions are reflected in the choice of hyperparameters and interpretation method, the Bayesian framework encodes assumptions via prior distributions and modeling choices. Additionally, the frequentist sliding-window approach quantifies evidence against a null hypothesis using p-values, whereas the Bayesian Wishart process approach provides uncertainty in its connectivity estimate, and uses this uncertainty to quantify evidence for both dynamic and static connectivity using Bayes factors.

Using simulations with different latent covariance structures, we evaluated both approaches. Overall, our simulations showed that both the Wishart process and the sliding-window method accurately estimate different types of connectivity, but the quality depends heavily on modeling assumptions. For example, when we use large window sizes for the sliding window approach or use smooth kernels such as the rational quadratic kernel in the Wishart process, we were unable to model fast fluctuations in connectivity. Our hypothesis testing results showed that these modeling choices also impact the test accuracies, with performance decreasing if the covariance itself was not captured accurately. For the Wishart process framework, this sensitivity was driven primarily by the choice of kernel rather than by the priors placed on the kernel hyperparameters. Overall, our robustness analysis showed relatively stable results across different prior settings for the kernel hyperparameters. Moreover, our simulations indicated that the outcomes are strongly influenced by choice of test statistic. While the variance and maximum power showed similar and reliable performances in terms of accuracy, recall, and false positive rates, median-crossing performed much worse. Compared to the other test statistics, median-crossing showed a reduced ability to detect dynamics, while also estimating many false positives for the exponential and periodic-exponential kernels. 

Our results for the HCP n-back working memory task fMRI dataset demonstrated how the two frameworks can be used to study dynamic functional connectivity in an empirical setting. In the realistic scenario that all connections show dynamics to some degree, the story becomes more difficult. Therefore, we focused on distinguishing between specific connections based on their dynamics, rather than classifying whether these connections were static or dynamic. In general, similar to our findings in the simulations, both testing frameworks were largely impacted by modeling choices and test statistics. We selected three regions of interest to compare the two frameworks. However, hypothesis testing of specific connections of interest on the individual level is challenging because of variations in functional connectivity between individuals~\cite{aguirre1998variability, mueller2013individual}. Results of our newly introduced pooled Wishart process model and the sliding-window group level estimates resulted in even more robust estimates than the individual-level estimates, and were in line with our expectations, showing that the connections between the DLPFC and IPL was modulated more strongly by working memory load than the connection between these areas and A1. 

In general, both the simulation results and empirical application highlighted, for both frameworks, the effect that test statistic and modeling choices can have on the conclusion that is drawn about dynamic functional connectivity. In practice, selecting the right kernel or window size before estimating functional connectivity over time is challenging. If prior knowledge about the expected structure of connectivity over time is known, this information can be used to choose a kernel or window size. However, when little is known about the expected structure, future work could instead focus on how to automatically learn a kernel or window size. For example, Duvenaud et al.~\cite{duvenaud2013structure} showed how to learn Gaussian process kernels from the data. Additionally, this could be combined with Bayesian model averaging, in which different kernel choices are combined, effectively learning a weighted combination of different kernels~\cite{hinne2020conceptual}. These approaches do suffer from issues with scalability when they are combined with the Wishart process. Resolving those issues is an important challenge for future work. Another interesting approach might be to combine Wishart processes with Gaussian process change-point detection methods~\cite{saatcci2010gaussian}, which would be a way to capture both continuous dynamics as well as sudden state-switches in functional connectivity.

In specific instances, we observed that with one kernel we found evidence for dynamic connectivity while for another kernel on the same dataset, we observed evidence for static connectivity. This illustrates that we are in fact not testing for the general existence of dynamics, but a specific type of dynamics that corresponds to the kernel that was specified beforehand. Although the current Bayesian hypothesis testing framework only distinguishes between static or dynamic functional connectivity over time, we could compare different prior assumptions using model comparison, for example by using the marginal likelihood estimated by the Sequential Monte Carlo sampler. This would provide us with more information about the type of dynamics.

There are several limitations that need to be taken into consideration. First, the Bayesian Wishart process framework makes use of the Savage-Dickey density ratio to compute the Bayes factors. This density ratio can only be used if the prior of the test-relevant parameters are independent of the prior of the nuisance parameters~\cite{wagenmakers_bayesian_2010, heck2019caveat}. In the context of our work, this means that the dynamic functional connectivity for a given connection should be independent of the {dynamic functional connectivity of all other pairs of brain regions. Since the estimated correlation matrices should be positive semi-definite, this constraint is violated when modeling more than two brain regions. To address this dependence between parameters, an alternative Savage-Dickey density ratio with a correction term has been proposed by Heck~\cite{heck2019caveat}, but this it is not straightforward to implement due to the high dimensionality of the Wishart process. Although incorporating this correction term would be a valuable improvement of the Bayesian framework, we have not applied this in the current work, as our focus was on comparing the different modeling assumptions of Bayesian and frequentist frameworks.

Another limitation is the computational scalability of inference of the Wishart process, here implemented using Sequential Monte Carlo sampling. For many practical purposes, the Wishart process might be prohibitively slow, since inference time of the Wishart process depends on the number of observations and regions, and on the convergence of the model parameters. Although more computationally efficient estimation procedures using variational inference have been proposed~\cite{heaukulani2019scalable}, these are less robust and risk failing to capture parts of the dynamics~\cite{huijsdens2024robust}, so these approaches should be used with care. In future work, improvements could be made here by combining sparse approximations of Gaussian processes~\cite{rossi2021sparse} and factorized covariance models~\cite{heaukulani2019scalable, rowe2002multivariate} with the Sequential Monte Carlo method. Additionally, although we have covered a wide range of potential dynamic connectivity structures, more elaborate model-based approaches to simulate more realistic fMRI data~\cite{wang2019inversion, deco2019brain}.  

Finally, it would be interesting for future work to extend the current Wishart process framework with a hierarchical Wishart process framework, in which not only a group-level connectivity estimates is inferred (like in our pooled Wishart process model), but also subject-specific connectivity estimates are modeled.

In summary, we have proposed an approach to hypothesis testing for dynamic functional connectivity using a Bayesian hypothesis testing framework combined with Wishart processes and compared it with the commonly used frequentist approach combined with the sliding-window method. While the frequentist sliding-window method provides evidence against a null hypothesis via p-values, the Bayesian framework quantifies evidence for both static and dynamic functional connectivity through Bayes factors, by also providing uncertainty over the estimated connectivity. However, we did not assess the calibration of these uncertainty estimates or whether this uncertainty is neurophysiologically meaningful. Future research should focus on making it easier and more transparent to determine which modeling assumptions should be chosen to model and test for dynamic functional connectivity~\cite{duvenaud2013structure}, especially when there is limited information about the dynamic structure in the functional connectivity of the data. One potential direction for this is the use of Bayesian model averaging to average over different kernels, although for the Wishart process this would first require improving the scalability of the inference of the model. This could help researchers to more robustly estimate and draw conclusions about dynamics in functional brain connectivity.

\section*{Funding}
Linda Geerligs was supported by a Vidi grant from the Dutch Research Council (NWO, VI.Vidi.201.150).

\section*{Acknowledgements}
We would like to thank Thomas Visser for his aid in preprocessing the Human Connectome Project working memory dataset.

\newpage
\bibliographystyle{ieeetr}

\newpage
\setcounter{figure}{0}    
\renewcommand{\thefigure}{S\arabic{figure}}
\renewcommand{\thesection}{S\arabic{section}}
\setcounter{section}{0}

\section*{Supplementary Materials}
\label{section:supplementary}

\section{Dynamic connectivity estimation methods}
Here we provide the formal description of the sliding-window method that was described in Section~\ref{subsubsection:sliding_window}, and the Wishart process that was described in Section~\ref{subsubsection:wishart_process}. 

\subsection{Sliding-window dynamic connectivity estimation} \label{supplementary:sliding_window}
For every input location $n=1,\ldots,\Observations$, we determine the starting point $l$ and end point $u$ of the window as $l=\min\left(\lfloor x_n-\frac{\windowlength-1}{2}\rfloor, 1\right)$ and $u=\max\left(\lfloor x_n+\frac{\windowlength-1}{2}\rfloor, \Observations\right)$, and with these we create a subset of observations $\v{S}_n= \left(\v{y}_l, \ldots, \v{y}_u\right)^\top$. For every new window, the input is shifted by $\tau$. With a stride length of 1, the total number of windows would therefore equal the number of observations, and larger stride lengths reduce the number of windows. Next, for each window, the pairwise covariance is computed as $\lfloor (\Observations - \windowlength)/\tau \rfloor + 1$
\begin{equation}
    \label{eq:sliding_window}
    \v{\Sigma}\left(x_n\right) = \frac{1}{\windowlength-1} \v{S}_n^\top \v{S}_n \enspace, \quad n = 1, \ldots, \Observations \enspace.
\end{equation}
By computing the pairwise covariance in Eq.~\eqref{eq:sliding_window} for every subset, we end up with an estimate of covariance as a function of the input $x_n$.   

\subsection{The Wishart process} \label{supplementary:wishart_process}
As explained in Section~\ref{subsubsection:wishart_process}, the Wishart process is constructed from a scale matrix $\v{L}$ and Gaussian processes that are parameterized by kernel hyperparameters $\theta$. To complete the generative model of the Wishart process, we define priors on these two parameters:
\begin{align*}
    \v{L} &\sim p\left( \v{L} \right) & \\
    \theta &\sim p\left( \theta \right) & \\
    f_{ki} \left(x_n \right) &\sim \mathcal{GP} \left( \mu \left( \cdot \right), \kappa \left( \cdot, \cdot;\theta \right) \right)  &\quad n& = 1, \ldots \Observations \enspace, \quad k = 1, \ldots \Dof \enspace, \quad i = 1, \ldots, \Regions \\
    \v{\Sigma}\left( x_n \right) & = \sum_{k=1}^{\Dof} \v{L} \v{f}_k \left( x_n \right) \v{f}_k \left( x_n \right)^\top \v{L}^\top \enspace &\quad n& = 1, \ldots, \Observations \\ 
    \v{y}_n \mid x_n &\sim \mathcal{MVN}_\Regions \left( \textbf{0}, \v{\Sigma}\left(x_n \right) \right) \enspace &\quad n& = 1, \ldots , \Observations \enspace.
\end{align*} 

Inferring the Wishart process from data, that is, computing the posterior distribution $p\left(\v{\Sigma}\left( \v{x} \right) \mid \v{Y}\right)$, is substantially more challenging than for the sliding-window approach. We follow the Wishart process inference procedure as described in~\cite{huijsdens2024robust}. That is, we use Gibbs sampling to infer the posterior distributions for (the lower Cholesky decomposition of) the scale matrix, the kernel hyperparameters, and the Gaussian process samples, and then use this to construct the posterior distribution for the covariance. To sample the scale matrix and kernel hyperparameters, we use a Random Walk Metropolis Hastings sampler. Since the Gaussian process samples are highly correlated, these are sampled using elliptical slice sampling~\cite{murray2010elliptical}.

Since parameters of the Wishart process are highly correlated, we use a Sequential Monte Carlo sampler~\cite{del2006sequential, huijsdens2024robust} \added{for robust inference. This} sampler requires a choice for the number of particles to be initialized, and the number of MCMC mutation steps to use. We initialized $1000$ particles. The number of MCMC mutation steps varied and was based on convergence of the posterior $p\left(\v{\Sigma} \left( \v{x} \right) \mid \v{Y} \right)$, as measured by the Potential Scale Reduction Factor ($\hat{R}$)~\cite{gelman1992inference}. We used three parallel inference chains, and set the number of mutation steps such that $\hat{R}<1.1$ for the covariance matrices. Although the kernel hyperparameters and scale matrix did not always converge across chains, the covariance itself did. This can happen because different combinations of the hyperparameters can result in the same covariance matrix, but since we are only looking at the covariance itself, our estimates of functional connectivity over time will be reliable. We implemented everything in Python. The Wishart process was implemented using the Python JAX framework~\cite{jax2018github} and SMC sampling in Blackjax~\cite{lao2022blackjax}, a Python library that builds on the JAX framework.

\section{The n-back task paradigm} \label{supplementary:working_memory_task}
Here we describe the working memory n-back task paradigm mentioned in Section~\ref{subsection:fmridata}. The task was a working memory n-back task in which subjects were shown a sequence of images and had to press a button if the current image matches the one shown $n$ steps back. The task alternated between blocks of 0-back and 2-back conditions, where the 0-back condition requires the participant to respond when the current stimulus matches a target image, and the 2-back condition asks the participant to respond when the current stimulus matches the one shown 2 steps back. Hence, the 2-back blocks require continuous updating of the sequence. The stimuli were images of either faces, places, tools, or body parts. A more detailed description of the task and data collection procedure is provided by Barch et al.~\cite{barch2013function}.

The dataset was preprocessed using the preprocessing pipeline as described by Glasser et al.~\cite{glasser2013minimal}. Additionally, we used high-pass filtering to remove any fluctuations slower than 0.008 Hz. The data contained 405 time points, which equals 291.6 seconds of fMRI recording. We removed the first five time points to reduce any potential artifacts related to the scanner equilibrium.

\section{Accuracy of connectivity estimates} \label{supplementary:all_mses}
To measure the accuracy of the estimated connectivity compared to the ground truth correlations, we compute the mean squared error (MSE). \deleted{However, to accurately reflect the differences between the Bayesian and frequentist frameworks, the MSEs are computed differently for the Wishart process and sliding-window method.} In the Bayesian framework\replaced{, we compute the MSE over the posterior mean.}{we make use of the full posterior distribution $p \left( \v \Sigma ( \v x ) \mid \v Y \right)$ when testing for dynamics. Therefore, the MSE is computed for each posterior sample and then averaged. This approach naturally results in larger MSEs, because the full posterior variability is incorporated into the MSE.} \deleted{In contrast, }\replaced{T}{t}he frequentist sliding-window method provides only single values instead of a full distribution, and therefore we compute the MSE directly on these. 

\begin{figure}[H]
    \centering
    \includegraphics[width=\linewidth]{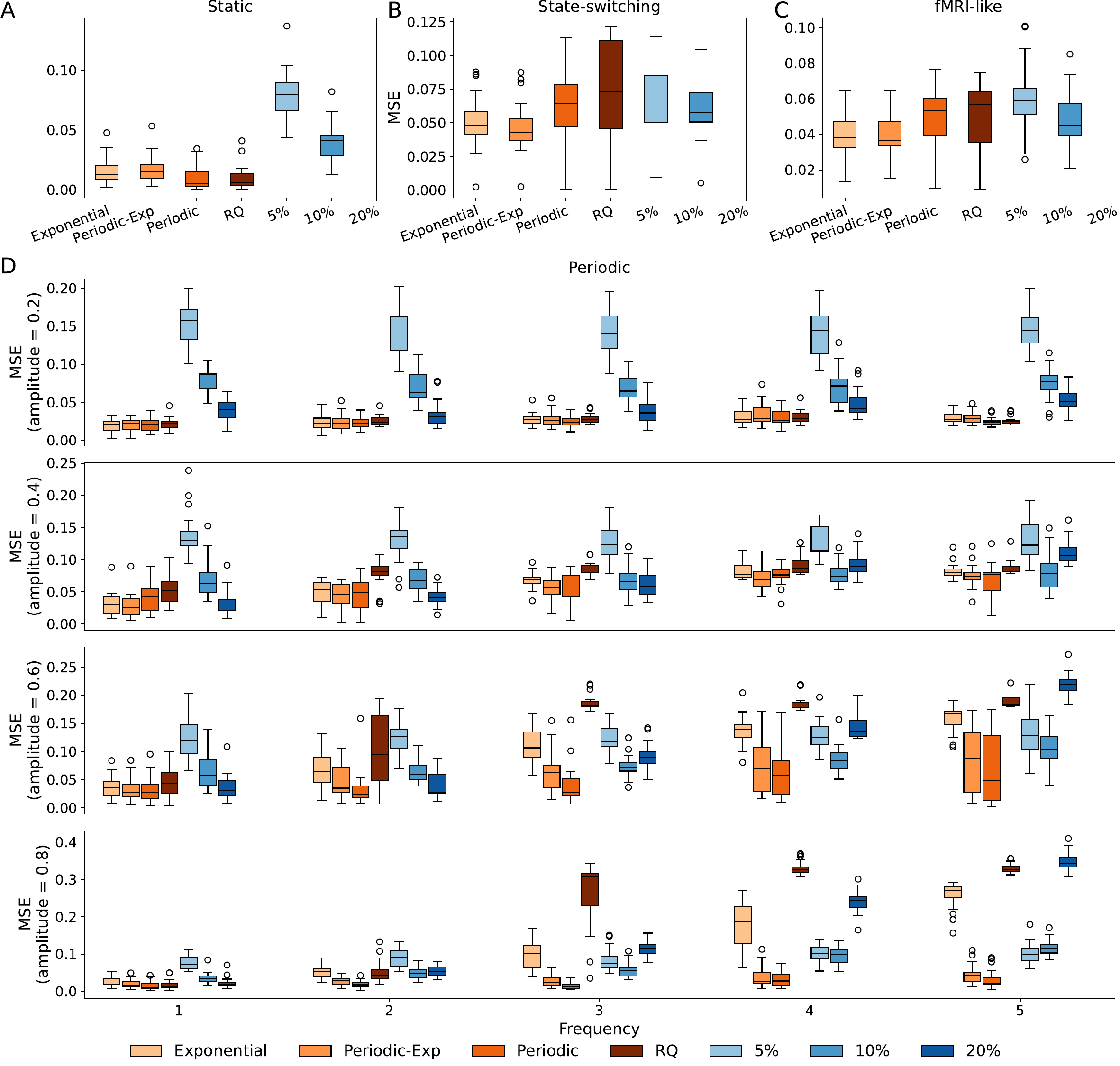}
    \caption{\textbf{Mean squared errors between true and estimated correlations for all simulations with 150 observations.} The figure shows the static (A), state-switching (B), fMRI-like (C) and periodic (D) simulations. For the Wishart process, the mean squared errors were computed over the posterior means.}
    \label{supplementary:mse_n150}
\end{figure}
\begin{figure}[H]
    \centering
    \includegraphics[width=\linewidth]{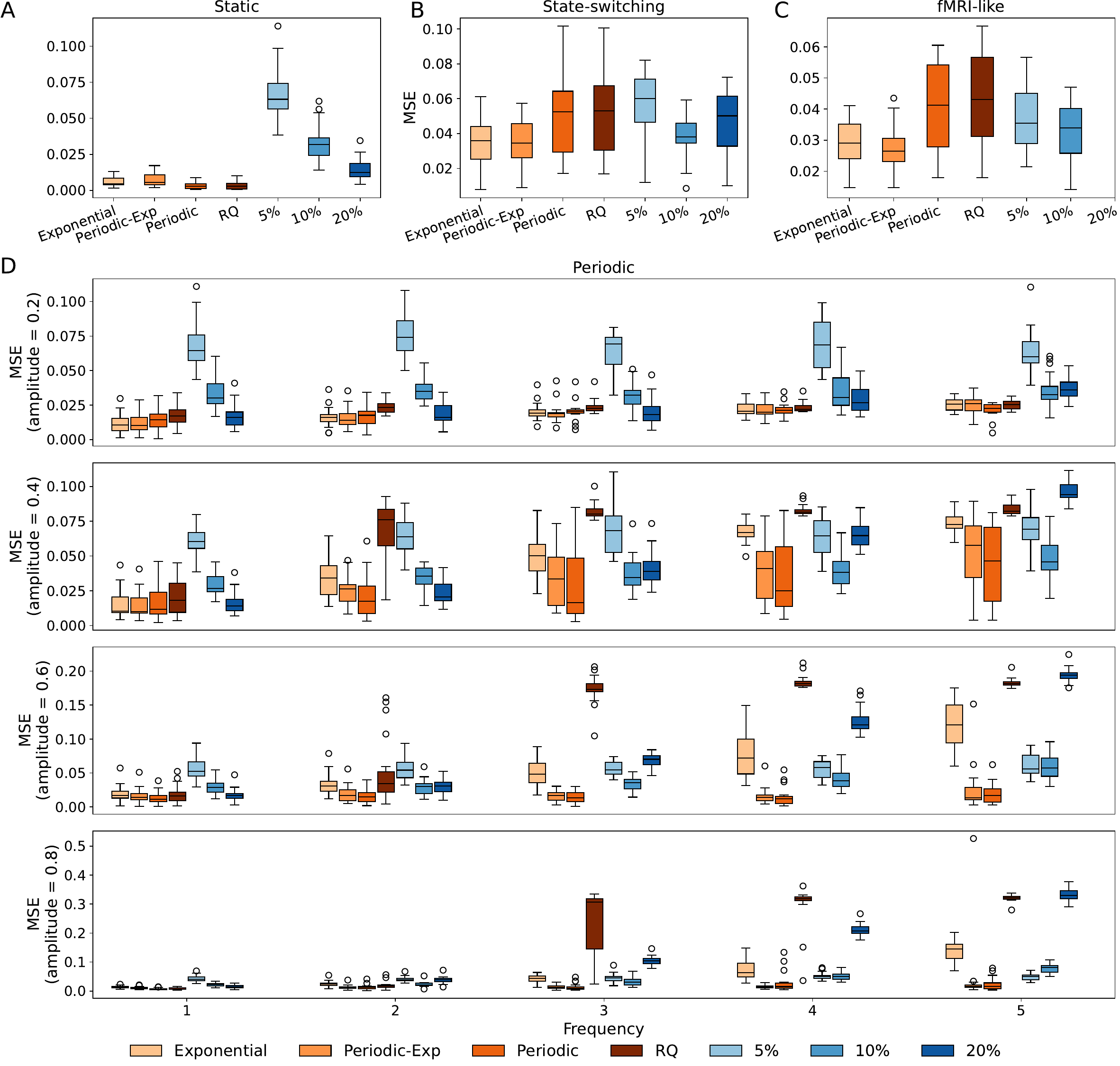}
    \caption{\textbf{Mean squared errors between true and estimated correlations for all simulations with 300 observations.} The figure shows the same simulations as in Figure~\ref{supplementary:mse_n150}, but with $\Observations=300$ instead of $\Observations=150$ observations.}
    \label{supplementary:mse_n300_mean_estimates}
\end{figure}
\begin{figure}[H]
    \centering
    \includegraphics[width=\linewidth]{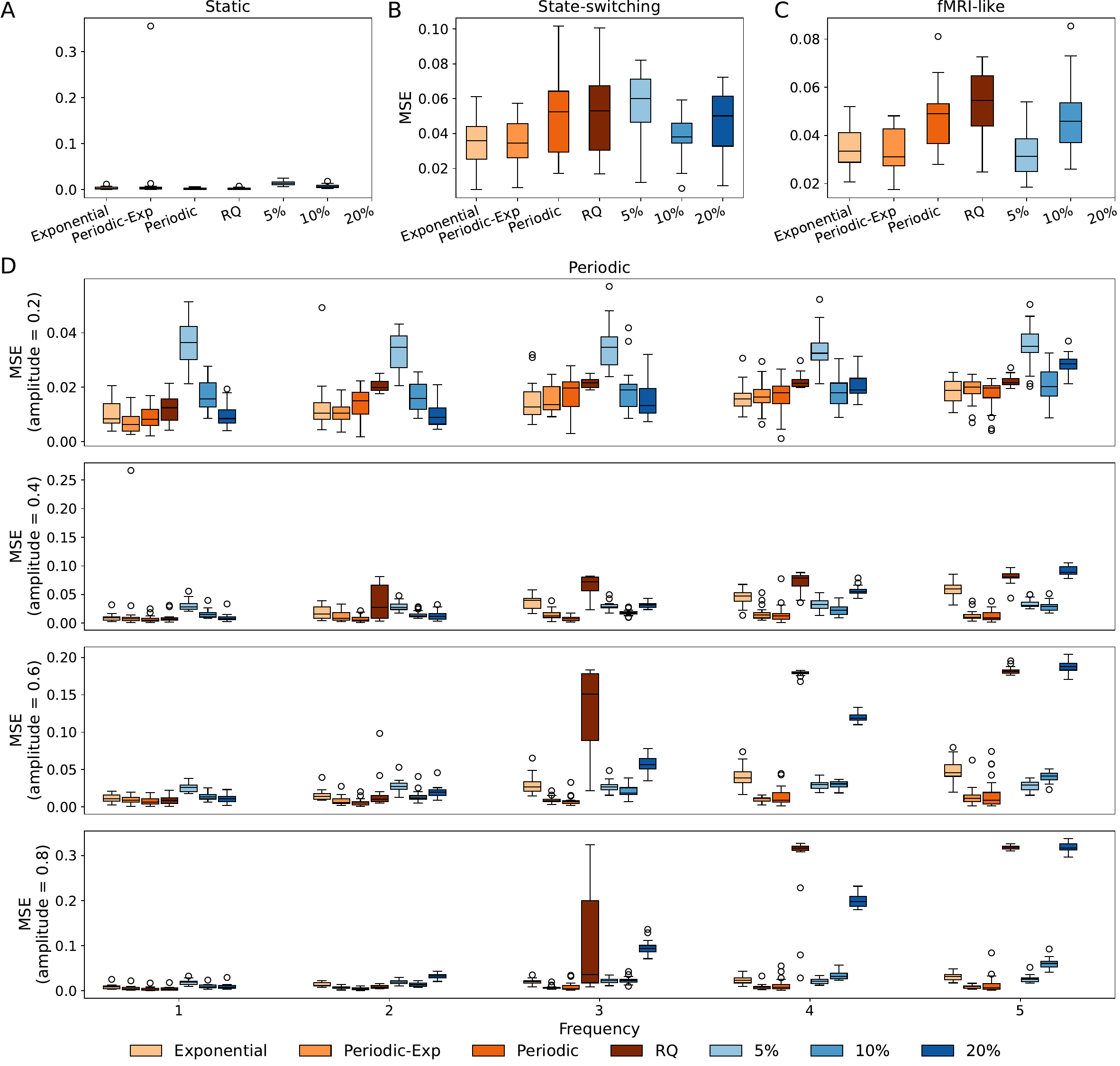}
    \caption{\textbf{Mean squared errors between true and estimated correlations for all simulations with 600 observations.} The figure shows the same simulations as in Figure~\ref{supplementary:mse_n150}, but with $\Observations=600$ instead of $\Observations=150$ observations.}
    \label{supplementary:mse_n600}
\end{figure}
In figures~\ref{supplementary:mse_n150}, ~\ref{supplementary:mse_n300_mean_estimates} and Figure~\ref{supplementary:mse_n600}, we provide all mean squared errors for $\Observations=150$, $\Observations=300$ and $\Observations=600$ observations. Overall, we observe similar patterns across different numbers of observations. Figure~\ref{supplementary:mse_n300_mean_estimates} presents the MSEs for $\Observations=300$ observations for the static, state-switching, fMRI-like and periodic simulations. Overall, the sliding-window method with a large window size (10\% and 20\% of the observations) and the Wishart process with a smooth kernel tend to recover static connections more accurately than dynamic connections, as could be seen in Figure~\ref{supplementary:mse_n300_mean_estimates}A. Figure~\ref{supplementary:mse_n300_mean_estimates}B presents the results from the state-switching simulations. Here the exponential\replaced{ and}{,} periodic-exponential\replaced{ kernels}{, and rational quadratic with a $\text{LogNormal}(-3, 1)$ prior set on $\alpha$} outperform the other kernels. The rational quadratic kernel \deleted{with $\text{LogNormal}(0, 1)$ priors on all hyperparameters (RQ1)} and the periodic kernel struggle with the state switches. The performance of the sliding-window method is highly dependent on window length. \added{The results for the fMRI-like simulation are shown in Figure~\ref{supplementary:mse_n300_mean_estimates}C. Again, the exponential and periodic-exponential kernels are better at modeling this covariance structure than the periodic and rational quadratic kernels.} In Figure~\ref{supplementary:mse_n300_mean_estimates}D, we observe that both the Wishart process with the exponential and rational quadratic kernels fail to capture correlations with higher frequencies. The sliding-window method with the largest window length also shows difficulties in modeling these faster dynamics. Finally, if the amplitude of the connectivity is decreased, we observe that all methods show increased difficulty in accurately modeling the connectivity with a lower amplitude, and there is no clear effect of frequency anymore.

\deleted{In figures~\ref{fig:mse_states_hrf_static} and \ref{supplementary:mse_n300}, the mean squared errors for the Wishart process estimates were computed over the entire posterior estimate, by first taking the mean squared error per sample and then averaging. To compare the performance with the sliding-window method and show the difference between the two metrics, Figure~\ref{supplementary:mse_n300_mean_estimates} presents the results for $\Observations=300$ observations when using only the posterior mean to compute the mean squared error. We observe that the MSE for the Wishart process, averaged over the MSEs of posterior samples, is larger than the MSE over the mean of the posterior. This is a standard effect of the Bayesian framework. Namely, as the posterior mean has a low MSE, individual samples that deviate from this mean will be less accurate. Importantly, this uncertainty can provide additional insights, such as when the Wishart process is very confident about its connectivity estimate, and when it is less certain.}

\section{Wishart process robustness analysis} \label{supplementary:robustness}
\added{In the main manuscript, we evaluated the effect of different kernel functions on the connectivity estimates and the resulting conclusions about dynamics. Here, we perform robustness checks to explore the sensitivity of the results to different prior choices for the kernel hyperparameters $\theta$, and the scale matrix $\v{V}$.}

\subsection{Prior assumptions for the kernel hyperparameters}
\begin{figure}[H]
    \centering
    \includegraphics[width=\linewidth]{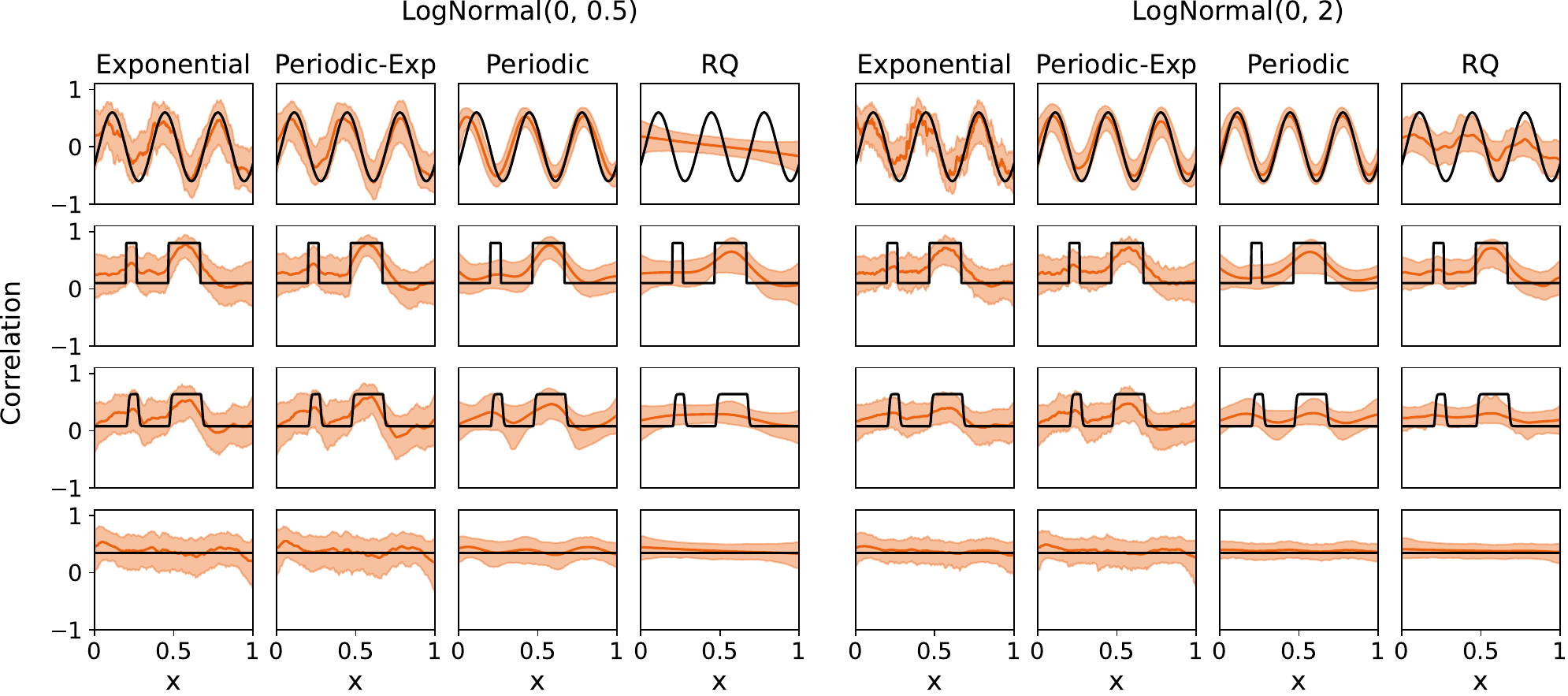}
    \caption{\textbf{Connectivity estimates with different prior assumptions on the kernel hyperparameters.} The figure shows the same simulations as in Figure~\ref{fig:connectivity_estimates}, but now with a narrower prior, $\text{LogNormal}(0, 0.5)$ (on the left), and a wider prior $\text{LogNormal}(0, 2)$ (on the right). Shown are the posterior mean and the 95\% highest density interval. The ground truth correlations are shown in black. We see that in most scenarios, the results are robust against changes to the prior.}
    \label{supplementary:priors_kernel_parameters}
\end{figure}
\added{The simulation in the main analysis assumed a log-normal prior with zero mean and standard deviation of one on all kernel hyperparameters. The inputs $\v{x}$ were scaled to the [0, 1] interval. Within this interval and with this prior specification, the kernels are allowed to model a wide range of covariance structures, ranging from very fast dynamics to almost constant connectivity.} 

\added{To assess robustness, we repeated the estimation procedure for the simulations from Figure~\ref{fig:connectivity_estimates} using a narrower log-normal prior, $\text{LogNormal}(0, 0.5)$, and a wider prior, $\text{LogNormal}(0, 2)$. Figure~\ref{supplementary:priors_kernel_parameters} shows the resulting connectivity estimates. For each combination of prior and simulation setup, we computed the mean squared error (MSE) between the posterior mean and the ground truth. Across most simulations, the connectivity estimates were very similar between the different priors. For all three priors, all MSEs with the ground truth were below 0.70, except for the rational quadratic kernel on the periodic simulation. This indicates that the estimated functional connectivity is largely robust to the priors on the kernel hyperparameters, with this single exception. Here, the broader $\text{LogNormal}(0, 2)$ prior performed better (MSE = 0.108) than the other two priors, which both had an MSE of 0.170. This suggests that a wider prior might be preferable for this specific simulation. For consistency, we decided to use the $\text{LogNormal}(0,1)$ prior throughout the analyses in the main text.}

\begin{figure}[H]
    \centering
    \includegraphics[width=\linewidth]{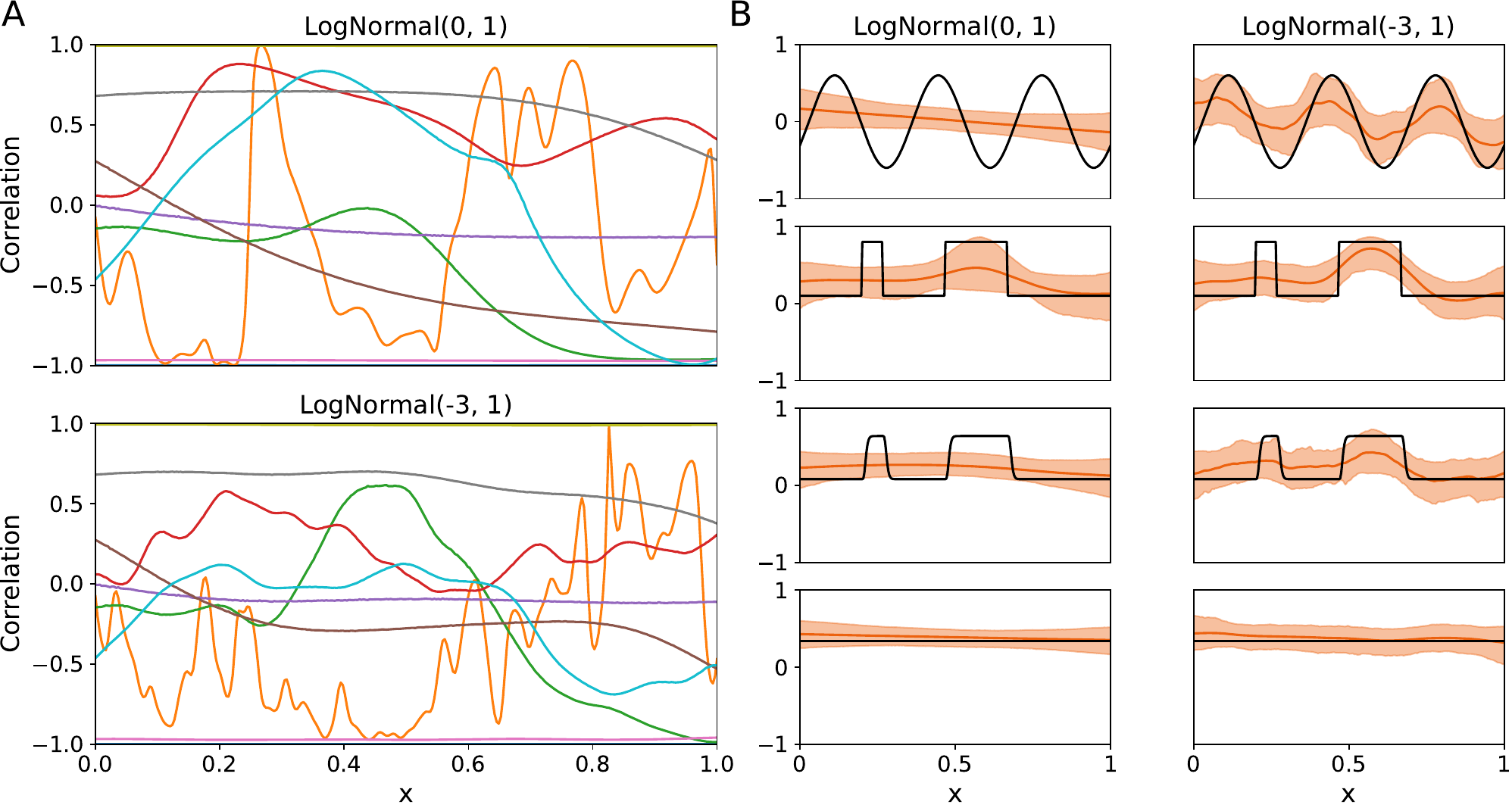}
    \caption{\textbf{Prior and posterior connectivity estimates with the rational quadratic kernel.} The figure shows prior samples of the connectivity based on two different priors for $\alpha$ (A) and the posterior estimates using these two priors for $\alpha$ on different simulations (B).}
    \label{supplementary:priors_RQ_min3}
\end{figure}
\added{For the rational quadratic kernel, we additionally evaluated another prior on the shape parameter $\alpha$. This parameter controls how strongly different lengthscales are mixed in the kernel. If $\alpha$ is very large, the rational quadratic kernel approximates the squared exponential kernel, and there is only one dominant lengthscale parameter. However, if $\alpha$ is small, multiple lengthscales are mixed in the kernel, allowing for both smooth trends and fast fluctuations simultaneously.}

\added{To evaluate robustness to the scale parameter, we considered instead $\alpha \sim \text{LogNormal}(-3, 1)$, while keeping the other kernel hyperparameter priors at $\text{LogNormal}(0, 1)$ as before. Figure~\ref{supplementary:priors_RQ_min3}A shows the effect of different priors on $\alpha$. The $\text{LogNormal}(-3, 1)$ prior for $\alpha$ has a higher density at small values of $\alpha$ and therefore results in connectivity that can alternate between smooth and quickly fluctuating. The posterior connectivity estimates on the four different simulations from Figure~\ref{fig:connectivity_estimates} are shown in Figure~\ref{supplementary:priors_RQ_min3}B, indicating that the rational quadratic kernel can be improved by appropriately informed prior choices, in this case with a preference for smaller mixing parameters $\alpha$. This is also confirmed by the MSEs between the posterior estimates and ground truth across the two priors, with the $\text{LogNormal}(-3, 1)$ prior on $\alpha$ resulting in an MSE of 0.080 for the periodic simulation, while the $\text{LogNormal}(0, 1)$ prior had an MSE of 0.170 on this simulation. All other MSEs were similar and below 0.07.}

\added{Overall, these results indicate that the connectivity estimates were largely robust across different prior specifications on the kernel hyperparameters. Across most simulations and priors, the posterior connectivity estimates remained similar. In some specific cases such as the rational quadratic kernel, more informative priors could improve estimation accuracy. However, specifying such informative priors on the kernel hyperparameters is only advised if there is any prior knowledge about the expected temporal structure of the functional connectivity available.}

\subsection{Prior assumptions for the scale matrix}
\begin{figure}[H]
    \centering
    \includegraphics[width=\linewidth]{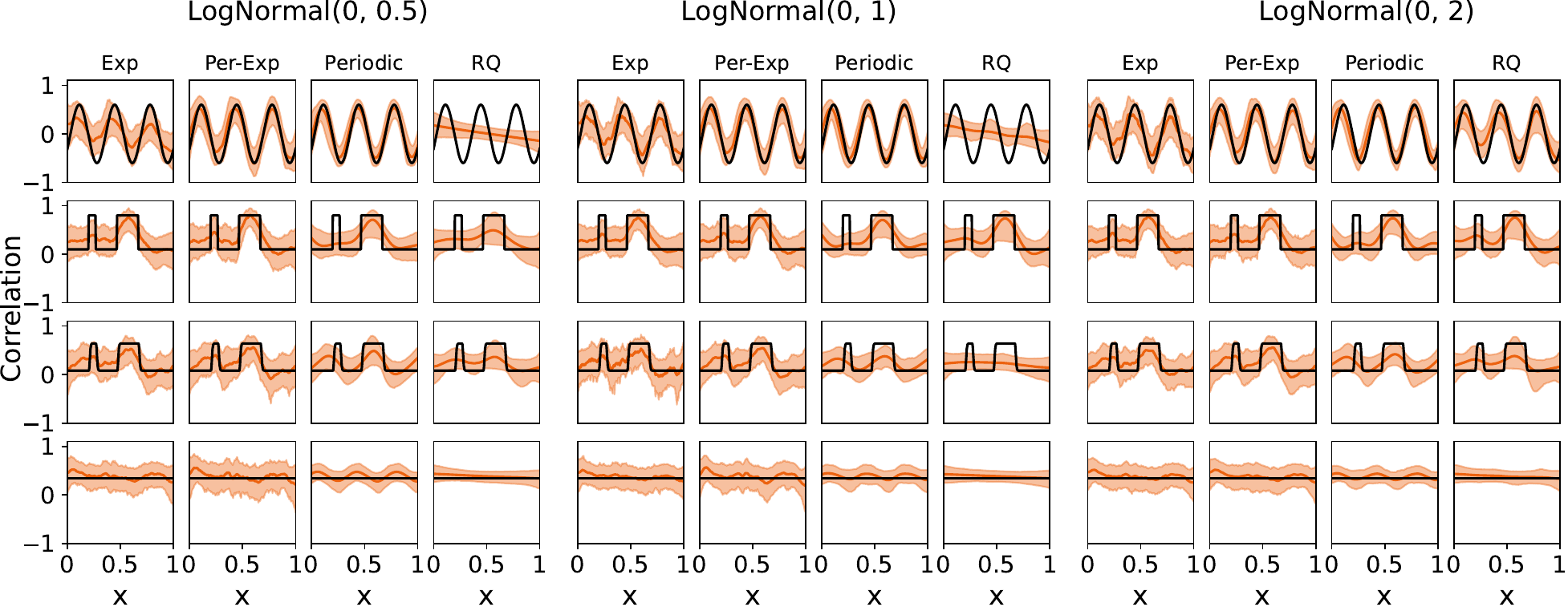}
    \caption{\textbf{Connectivity estimates with different prior assumptions on the scale matrix $\v{V}$.} The simulations that are shown are the same as in Figure~\ref{fig:connectivity_estimates}, but here they were modeled with an Lewandowski-Kurowicka-Joe prior on the scale matrix. The variance term in this prior was altered. We show results for values of $\sigma^2$ of 0.5, 1 and 2. The ground truth correlations are shown in black.}
    \label{supplementary:priors_scale_matrix}
\end{figure}
\added{In the main analysis, we assigned standard normal priors to each element of the lower Cholesky decomposition of the scale matrix. To evaluate robustness of our results to different prior specifications, we compare our results with a Lewandowski-Kurowicka-Joe (LKJ) prior~\cite{lewandowski2009generating} for the scale matrix. The LKJ distribution provides an interpretable and flexible prior, because this prior allows us to separately alter the prior assumptions on correlation and variance, as follows:
\begin{align*}
    \v{L} &\sim \text{LKJ-Cholesky} \left( \eta \right) \enspace, \\
    \v{D}_{ii} &= \text{LogNormal}(0, \sigma^2) \quad i = 1, \ldots, \Regions \enspace, \\
    \v{V} &= \v{D}\v{L} \v{L}^\top \v{D}\enspace.
\end{align*} 
We set the shape parameter $\eta$ of the LKJ prior to 1, as this implies a uniform prior over all positive definite correlation matrices. To assess the effect of different scale matrix priors, we explored the results for different values of the variance parameter $\sigma^2$, namely $\sigma^2 \in \{ 0.5, 1, 2 \}$. Similar to the main analysis, the kernel hyperparameter priors were all kept fixed at $\text{LogNormal}(0, 1)$. Figure~\ref{supplementary:priors_scale_matrix} shows the estimates for the three different scale matrix priors. }

\added{To measure robustness, we again compute the MSEs between the posterior mean of the estimates and the ground truth. For $\sigma^2=0.5$ and $\sigma^2=1$, MSEs were highly consistent with those from the main manuscript. All MSEs were below 0.070 except for the rational quadratic kernel in the periodic simulation, which had an MSE of 0.171 for $\sigma^2=0.5$ and of 0.162 for $\sigma^2=1$. This indicates that the model is robust to the choice of scale matrix prior.}

\added{Interestingly, by increasing the variance of the prior on the variance terms, the rational quadratic kernel outperformed the periodic estimate from the main analysis, with an MSE of 0.033 compared to 0.170 in the main analysis. All other MSEs remained similar, namely again below 0.070.}

\added{Overall, the connectivity estimates were robust to both the prior specifications of the scale matrix and kernel hyperparameters, as connectivity estimates remained highly similar across prior choices. However, the results of the rational quadratic kernel suggest that we might benefit from broader and more uninformative priors on both the scale matrix and the kernel hyperparameters than the priors that were used in the main analysis, especially when modeling fast fluctuations. This is consistent with the previous robustness analysis that showed that different priors may be beneficial for a kernel as flexible as the rational quadratic.}

\section{Additional simulation study results} \label{supplementary:additional_simulations}
Here we present the results of simulations that were not presented in the main text. We follow the same structure as Section~\ref{section:simulation_results}, by first focusing on the hypothesis testing results and then on the uncertainty provided by the Bayesian framework. 

\subsection{Hypothesis test performances}
\begin{figure}[H]
    \centering
    \includegraphics[width=\linewidth]{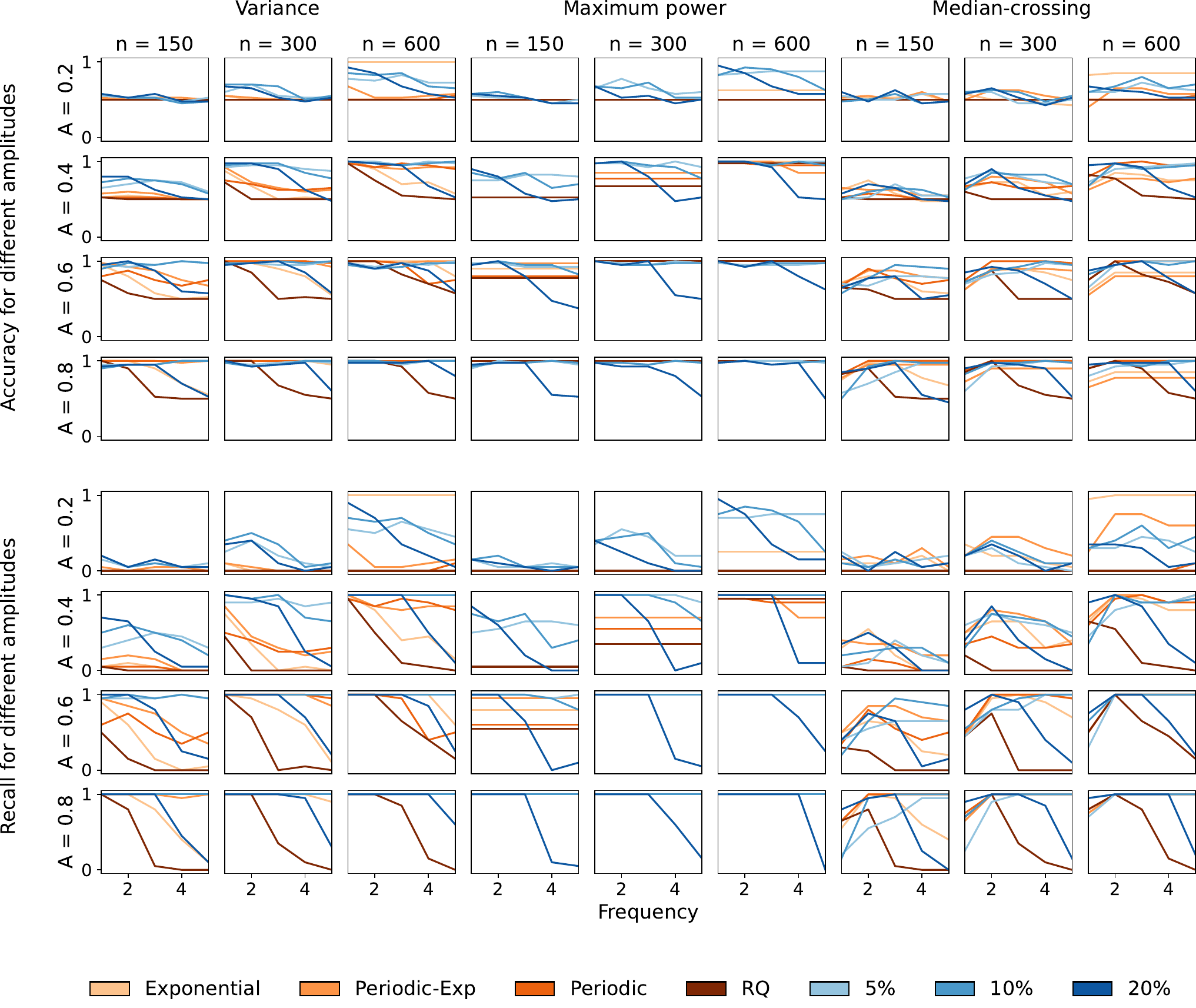}
    \caption{\textbf{Hypothesis test performances on all periodic simulations.} Performance for amplitudes of 0.2--0.8 are shown. The figure shows the fractions of correctly classified connections, out of all connections (accuracy) and out of only dynamic connections (recall) when using the variance, maximum power and median-crossing test statistics.}
    \label{supplementary:hypothesis_testing_n150_n600}
\end{figure}
\deleted{Figure~\ref{supplementary:hypothesis_testing_n150_n600} shows the accuracy and recall of all periodic simulation studies for all three statistics. These results show that the performance increases with more observations.}
\begin{figure}[H]
    \centering
    \includegraphics[width=\linewidth]{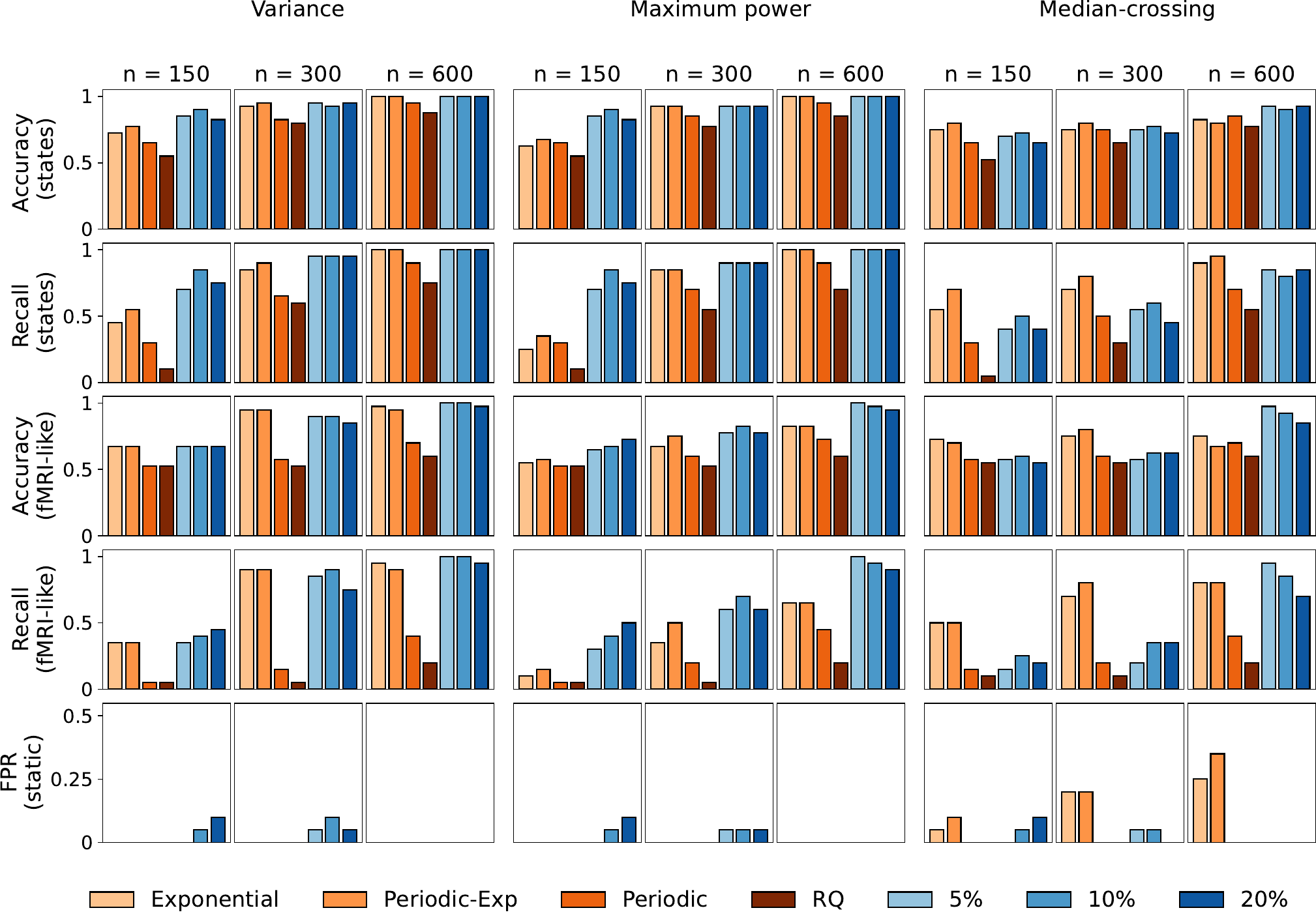}
    \caption{\textbf{Hypothesis test performances on the state-switching, fMRI-like and static simulations.} The figure shows accuracies and recalls (on the state-switching and fMRI-like simulations) and false positive rates (on the static simulations) of the variance, maximum power and median-crossing statistics.}
    \label{supplementary:results-hypothesis-testing-nonperiodic}
\end{figure}
\added{Figure~\ref{supplementary:hypothesis_testing_n150_n600} shows the accuracy and recall of all periodic simulation studies for all three statistics and} Figure~\ref{supplementary:results-hypothesis-testing-nonperiodic} presents the results for the non-periodic state-switching\added{, fMRI-like} and the static simulations. Here, the false positive rate indicates how many of the static simulations were incorrectly classified as dynamic. Overall, performance increases with more observations\deleted{, which is in line with the results on the periodic simulations}. 

The median-crossing and variance statistics show similar performance patterns between kernels and window sizes. However, the median-crossing statistic does not accurately detect slow dynamics, as can be observed by the relatively low recall scores at a frequency of 1. Both the variance and maximum power accurately detect dynamics of different frequencies for $\Observations=600$ observations, and $\Observations=300$ observations with an amplitude of 0.6 or 0.8, but performance decreases with $\Observations=300$ observations and an amplitude of 0.2 or 0.4. An interesting exception to this is the exponential kernel, which, with $\Observations=600$ observations, can accurately distinguish dynamics with an amplitude of 0.2 from static connectivity. Corresponding posterior variance distributions of this simulation are shown in Supplementary Figure~\ref{supplementary:posteriors02}.

\subsection{Uncertainty in Bayesian hypothesis testing}
\begin{figure}[H]
    \centering
    \includegraphics[width=\linewidth]{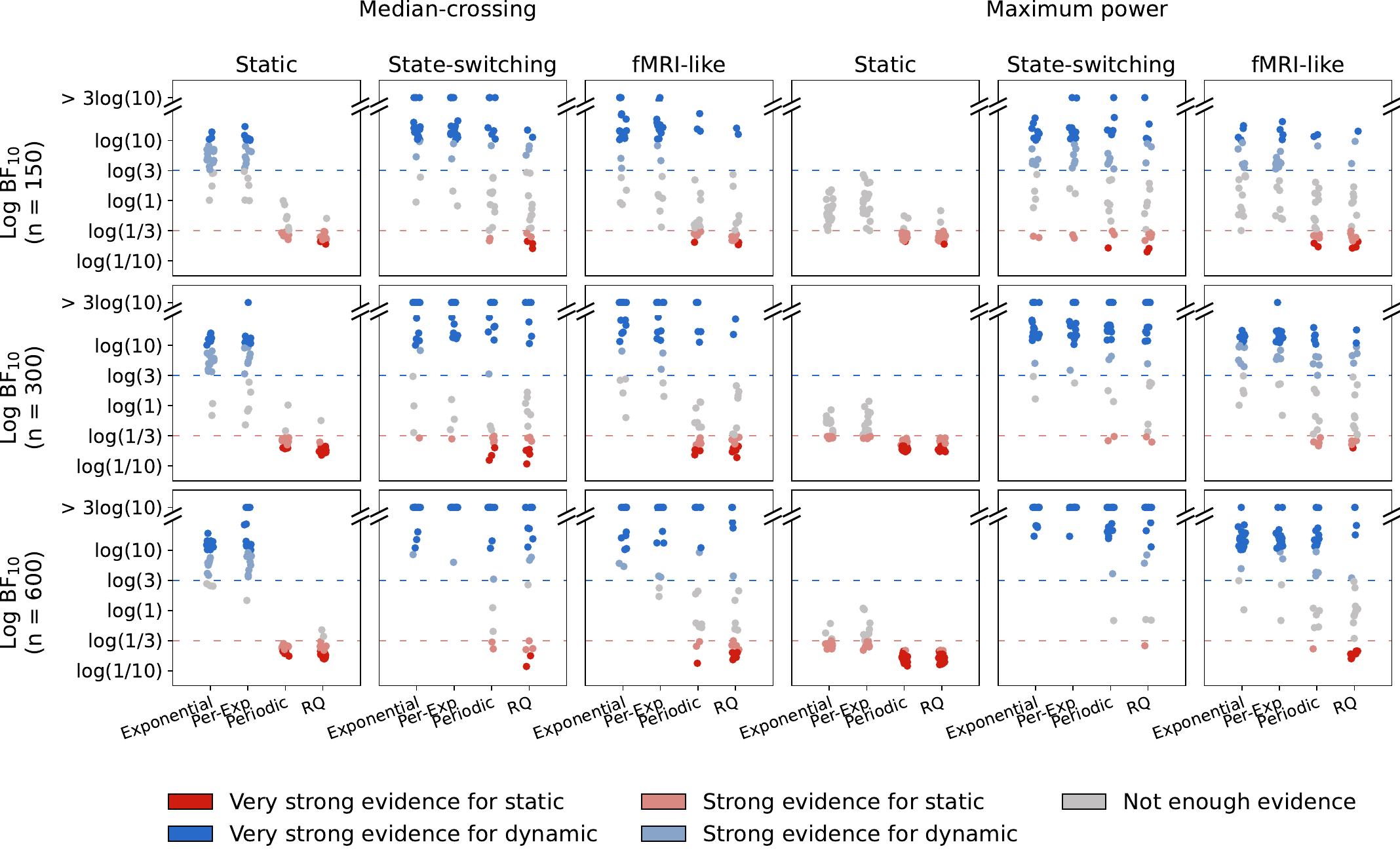}
    \caption{\textbf{Log Bayes factors of detecting dynamics for the static, state-switching, and fMRI-like simulations.} The figure shows the Bayes factors based on the median-crossing and maximum power test statistics. Every dot represents a single connection and is colored based on the amount of evidence of the connection being dynamic or static.}
    \label{supplementary:bayes_factors_mediancrossing}
\end{figure}
\added{In the main manuscript in Figure~\ref{fig:bayes-factors}, we only presented the Bayes factor distributions based on the variance test statistic. The results based on the median-crossing and maximum power test statistic are provided in Supplementary Figure~\ref{supplementary:bayes_factors_mediancrossing}.} Based on all three statistics, a consistent pattern is shown for the static simulations. Namely, the exponential and periodic-exponential kernels provide higher Bayes factors compared to the other kernels and these Bayes factors also clarify the large number of false positives that we observed in Figure~\ref{supplementary:results-hypothesis-testing-nonperiodic}.

\begin{figure}[H]
    \centering
    \includegraphics[width=\linewidth]{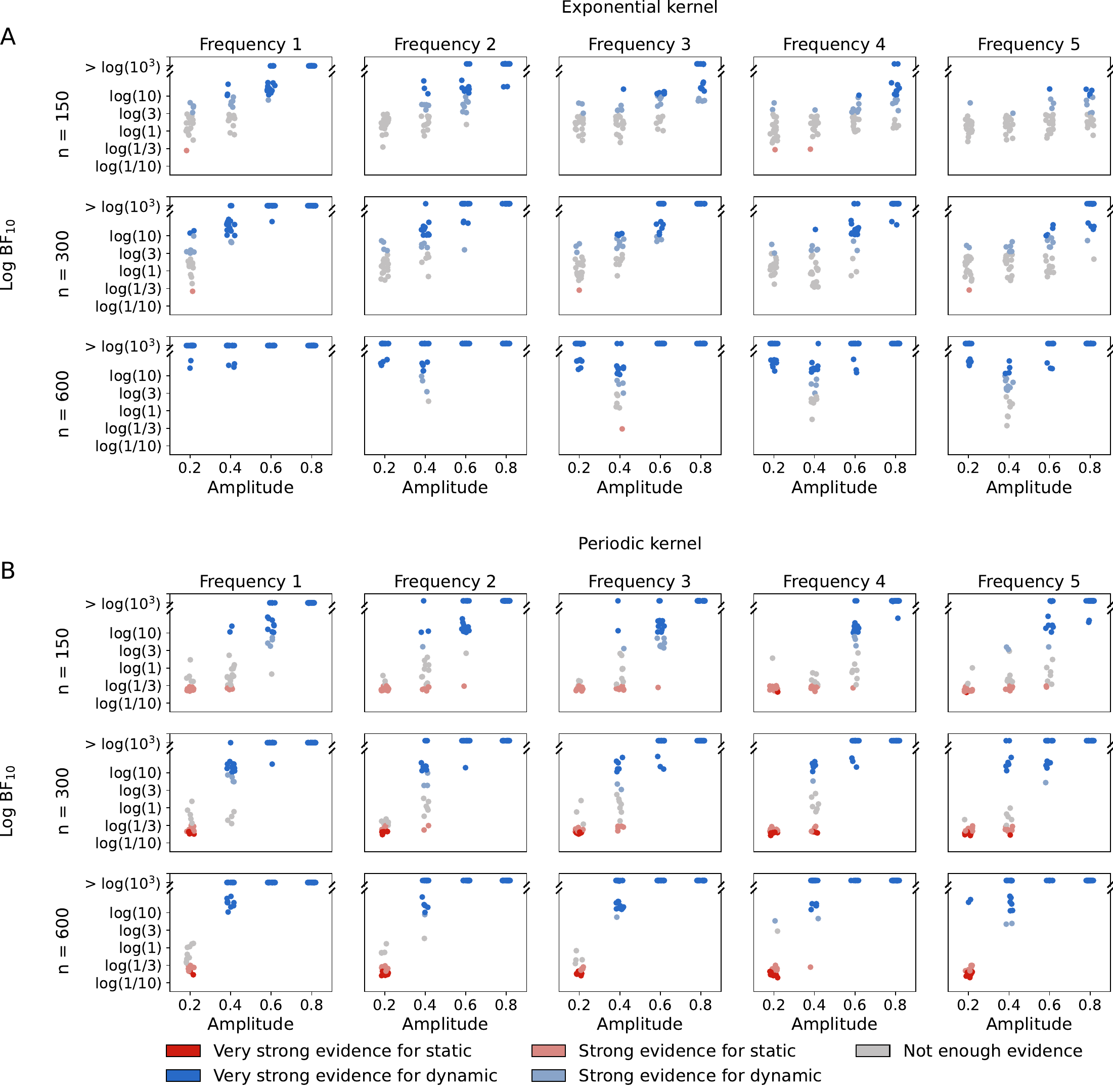}
    \caption{\textbf{Log Bayes factors of detecting dynamics for periodic simulations.} The figure shows the Bayes factors for the exponential (A) and periodic (B) kernels and are based on the variance test statistic. Every dot represents a single connection and is colored based on the amount of evidence of the connection being dynamic or static.}
    \label{supplementary:bayes_factors_periodic}
\end{figure}
Finally, Figure~\ref{supplementary:bayes_factors_periodic} presents the distribution of log Bayes factors for the simulations with a periodic covariance
structure, for $\Observations \in \{ 150, 300, 600\}$ observations, amplitudes of $0.2-0.8$, and frequencies of $1-5$. To show the effect of the choice of kernel on the log Bayes factors, we show the results of two distinct kernels here, namely the exponential and periodic functions. Each point in the figure represents a single connection and is colored based on the strength of its evidence. For both kernels, the Bayes factors indicate that the strength of evidence increases with the number of observations. For $\Observations = 150$ observations, a large number of edges is inconclusive. Moreover, the number of connections with conclusive evidence increases with an increasing amplitude and a decreasing frequency. In the case of the exponential kernel, there are more connections with inconclusive evidence with more rapid dynamics or dynamics with a lower amplitude, as these are estimated less accurately. For the periodic kernel, the strength of evidence hardly decreases with frequency.

\section{Additional empirical results} \label{supplementary:additional_empirical}
\begin{figure}[H]
    \centering
    \includegraphics[width=\linewidth]{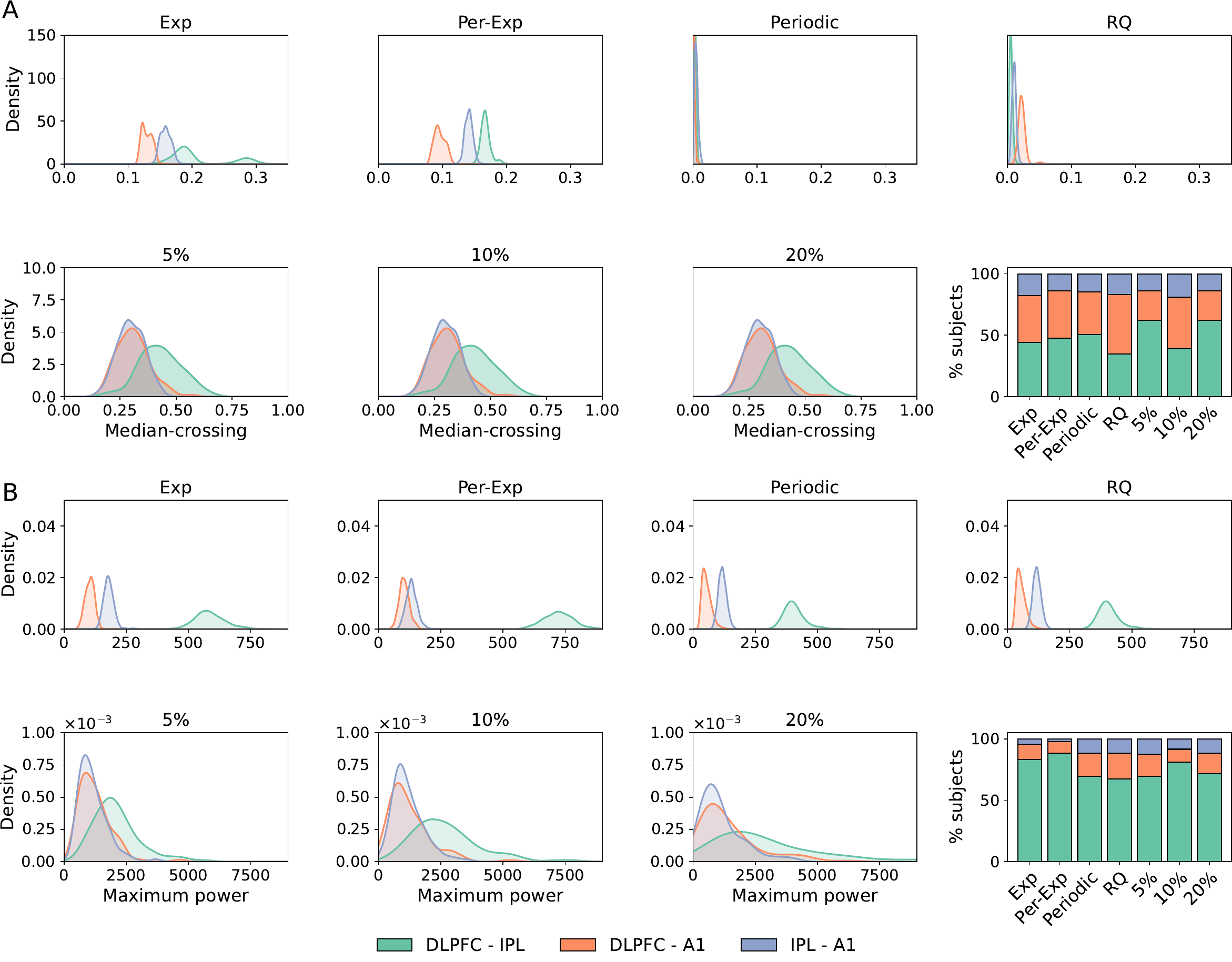}
    \caption{\textbf{Median-crossing (A) and maximum power (B) test statistics for three pairs of brain regions from the working memory task.} The barplots show the orderings of subject-level test statistics, whereas the density plots show the group-level test statistics.}
    \label{supplementary:empirical_other_test_statistics}
\end{figure}
\added{In the main manuscript, we presented the empirical results based on the variance test statistic. Here, we additionally show the posterior distributions and orderings based on the median-crossing (Figure~\ref{supplementary:empirical_other_test_statistics}A) and maximum power (Figure~\ref{supplementary:empirical_other_test_statistics}B) test statistics. The findings of these test statistics are similar to those in the main manuscript. However, differences between the connections are less pronounced when using the median-crossing test statistic, showing the importance of carefully selecting a test statistic.}

\section{Effect of different prior assumptions on the evidence} \label{supplementary:effect_different_priors_bayesfactor}
\begin{figure}[H]
    \centering
    \includegraphics[width=\linewidth]{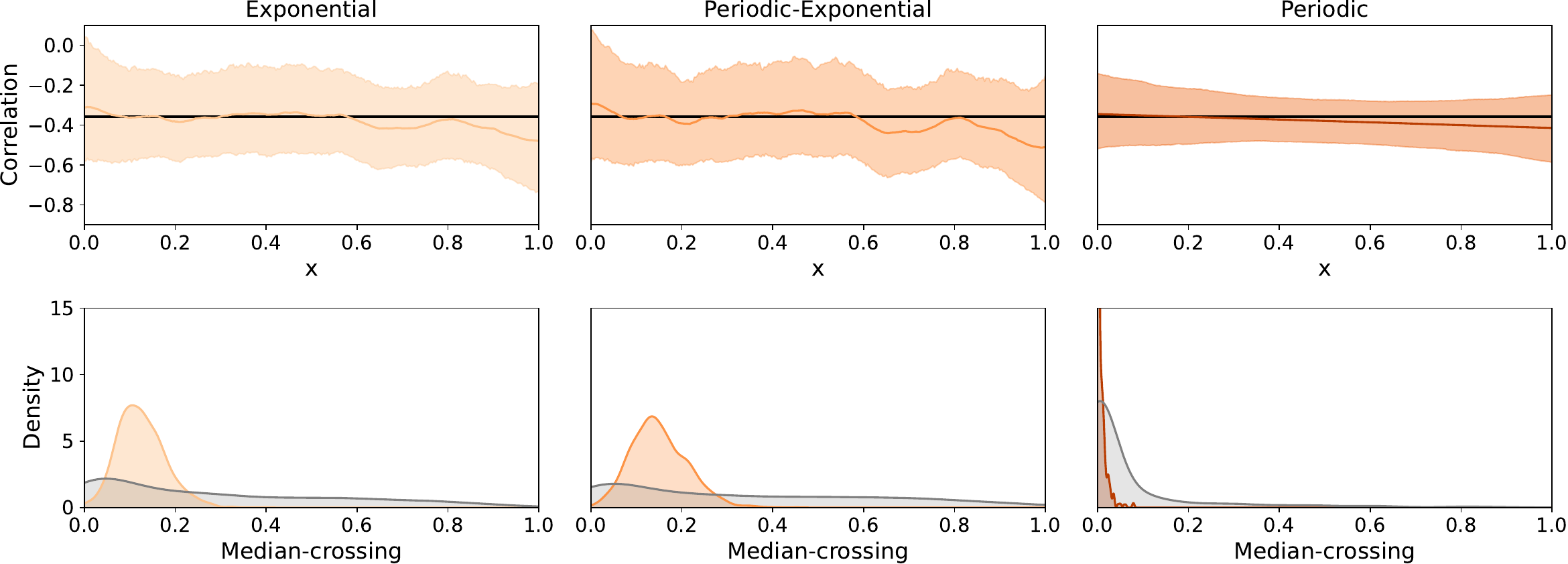}
    \caption{\textbf{Static connectivity estimates by the Wishart process and their corresponding prior and posterior median-crossing distributions.} Estimates using different kernels are shown. Prior distributions are shown in gray. The black line in the upper plots indicates the true latent connectivity.}
    \label{supplementary:priors_and_posteriors}
\end{figure}
Figure~\ref{supplementary:priors_and_posteriors} provides an explanation for the differences in Bayes factors that we observed, especially by the exponential and periodic-exponential kernel. This figure shows examples of connectivity estimates and its corresponding prior (in gray) and posterior distributions over the median-crossing statistic. All examples are based on a static true connectivity. For the periodic kernel, the prior probability of the test statistic being zero is much larger than for the exponential and periodic-exponential kernels, indicating a stronger prior bias towards static connectivity. In contrast, the exponential and periodic-exponential kernels have a relatively low prior probability of the test statistic being zero, indicating that these kernels model more dynamic correlations by design. This behavior is also observed in the prior covariance samples in Figure~\ref{fig:null-distributions-for-different-kernels}, where the exponential and periodic-exponential kernels produce samples that contain many small fluctuations compared to the relatively smooth periodic kernel. These differences in prior distributions directly influence the resulting log Bayes factors through the Savage-Dickey density ratio (described in \eqref{eq:savage_dickey}), as the Savage-Dickey density ratio compares the prior and posterior densities at the null value, which in our case is at a value of zero. As a result, even if two kernels would provide nearly identical posterior distributions over a test statistic, their log Bayes factors can substantially differ because of their differences in priors. For example, if all three kernels would find a posterior test statistic centered far above zero (suggesting dynamic connectivity), the Bayes factor for the periodic kernel would be stronger because it has a larger density at zero. 

\begin{figure}[H]
    \centering
    \includegraphics[width=\linewidth]{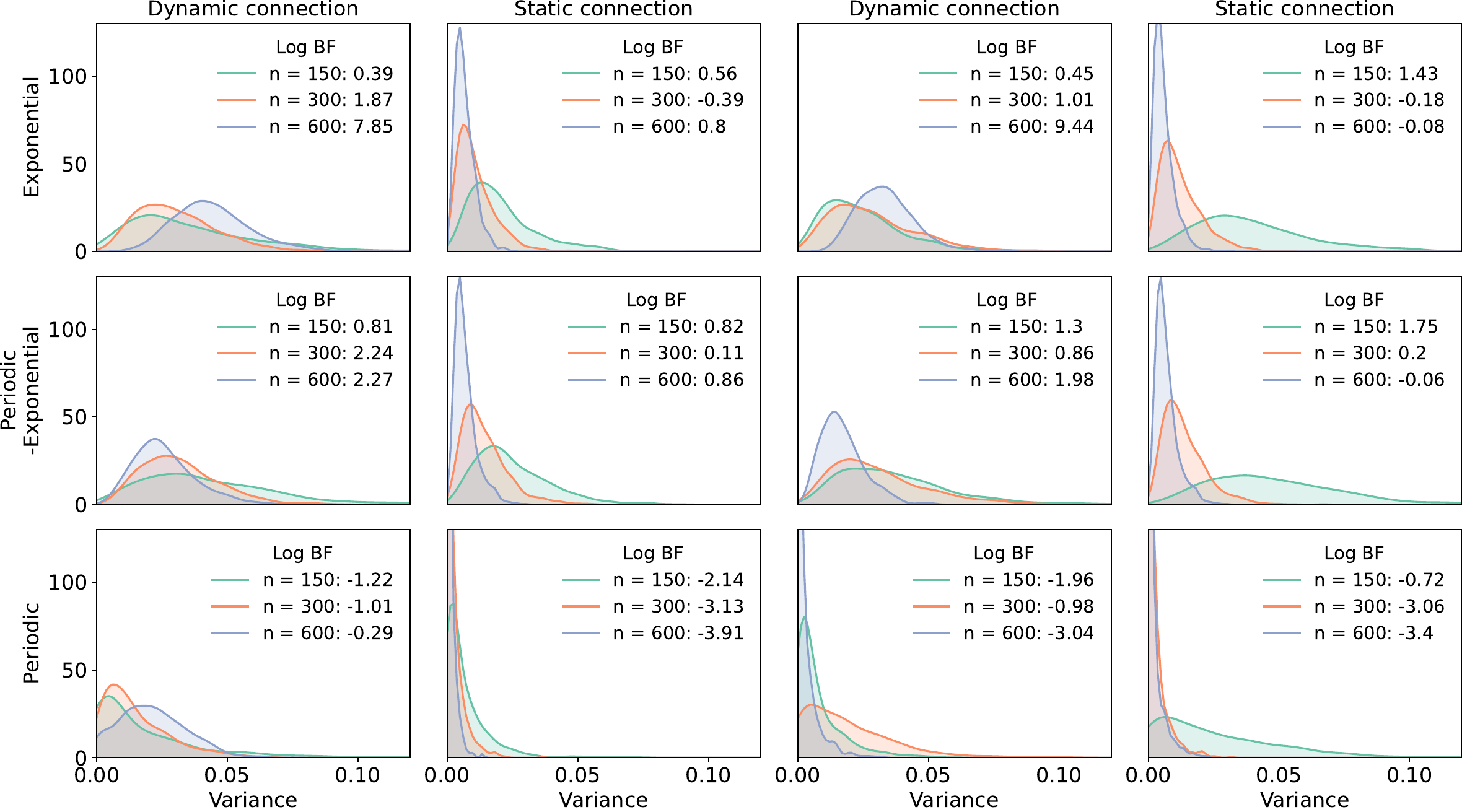}
    \caption{\textbf{Posterior variance distributions for different numbers of observations and different kernel choices.} The figure presents the results of four different connections, of which two are static and two are dynamic with a frequency of 2 and an amplitude of 0.2.}
    \label{supplementary:posteriors02}
\end{figure}
Moreover, Figure~\ref{supplementary:posteriors02} shows the posterior variance distributions and corresponding log Bayes factors for four  connections, of which two were simulated static and two dynamic with a frequency of 2 and an amplitude of 0.2. This figure shows that especially the exponential kernel is able to capture dynamics with a small amplitude, as indicated by the relatively low density at a variance of 0.

\end{document}